\documentclass[preprint,12pt,sort&compress]{elsarticle}

\usepackage{titletoc}
\usepackage[T1]{fontenc}
\usepackage{amssymb}
\usepackage{amsmath}
\usepackage{array}
\usepackage{adjustbox}
\usepackage{textcomp}
\usepackage[utf8]{inputenc}
\usepackage{blindtext}
\usepackage{xspace}
\usepackage{hologo}
\usepackage{geometry}
\usepackage{lscape}
\usepackage{csquotes}
\usepackage{multirow}
\usepackage{subcaption}  
\usepackage{makecell}
\usepackage{array}
\usepackage{hyperref}
\usepackage{tabularx}
\usepackage{xcolor}
\journal{Solar Energy Materials and Solar Cells}

\begin{document}

\begin{frontmatter}




\title{Hierarchically structured, high-efficiency and thermally robust perovskite solar cells with band-engineered double-hole layers}

\author[inst1]{Md. Faiaad Rahman}
\author[inst1]{Arpan Sur}
\author[inst1]{Ahmed Zubair\corref{cor1}}
\ead{ahmedzubair@eee.buet.ac.bd}
\cortext[cor1]{Corresponding author:}

\affiliation[inst1]{organization={Department of Electrical and Electronic Engineering}, 
            addressline={Bangladesh University of Engineering and Technology}, 
            city={Dhaka}, 
            postcode={Dhaka-1205}, 
            country={Bangladesh}}

\begin{abstract}

Despite the promising optoelectronic properties of methylammonium lead iodide (MAPbI\textsubscript{3})-based perovskite solar cells (PSCs), their commercial viability is hindered by interfacial energy misalignment, suboptimal light absorption, and thermal instability. Here, we present a comprehensive theoretical framework to enhance the power conversion efficiency (PCE) of MAPbI\textsubscript{3} based PSC through integrated electronic and morphological engineering. Firstly, to address interfacial recombination and inefficient hole extraction at the perovskite/HTL junction, we introduced a double hole transport layer (HTL) stack comprising CuO and I\textsubscript{2}O\textsubscript{5}-doped Spiro-OMeTAD, which significantly improved energy level alignment and carrier selectivity. Comprehensive multiphysics simulations, combining finite-difference time-domain (FDTD) optical analysis with finite element method (FEM) based electrical and thermal modeling, demonstrated that optimized doping concentrations and thickness parameters within the CuO/Spiro-OMeTAD hole transport layers can enhance the PCE to 22.78\%. However, planar architectures, while offering ease of fabrication and scalability, exhibit weak near-UV and near-infrared absorption, whereas nanostructures attain superior light trapping but incur significant fabrication complexity, underscoring the need for balanced design strategies.
To address these inherent limitations, we propose a hierarchical ellipsoidal patterned solar cell (HEPSC), wherein a top-layer ellipsoidal nanostructure is introduced across the full device stack. This design enhances broadband light trapping and optical confinement throughout the active layers while maintaining fabrication feasibility through geometrically realistic structuring. The optimized HEPSC achieved a maximum PCE of 26.39\%, with short-circuit current density, open-circuit voltage, and fill factor reaching 29.29 mA/cm\textsuperscript{2}, 1.074 V, and 83.87\%, respectively, under isothermal steady-state conditions. Finally, to assess the thermalization effect in the proposed design, a coupled opto-electro-thermal simulation reveals that the HEPSC retains 94.5\% of its efficiency under non-isothermal conditions (up to 52$^\circ$C). Collectively, these strategies provide an integrated pathway for designing efficient, morphologically optimized, and thermally resilient next-generation thin-film PSCs.
\end{abstract}

\end{frontmatter}



\section{Introduction}
The escalating global energy demand and pressing environmental concerns underscore the critical need for sustainable and efficient renewable energy technologies. Among these, photovoltaics (PV) technologies stand out for their capacity to provide clean, direct solar-to-electric energy conversion. For several decades, silicon-based solar cells have dominated the PV market, with commercial solar cells surpassing 24\% PCE and benefiting from a well-established manufacturing ecosystem \cite{Liang2025}. However, despite their maturity, silicon solar cells face several inherent limitations. These include high energy and capital costs associated with crystal growth and wafer processing, rigid and bulky module designs that hinder flexible or lightweight applications, and diminishing returns in efficiency due to their proximity to the Shockley–Queisser limit of ~29.4\% for single-junction cells~\cite{andreani2019silicon}. Moreover, silicon's indirect bandgap necessitates thicker absorber layers, increasing material consumption and limiting transparency, which restricts applications in building-integrated photovoltaics (BIPV) and tandem configurations. Perovskite solar cells (PSCs), based on hybrid organic-inorganic halide compounds with a general formula ABX\textsubscript{3}, have rapidly emerged as one of the most promising next-generation photovoltaic technologies~\cite{afroz2025perovskite}. PSCs offer a compelling combination of high efficiency, low fabrication cost, and ease of processing, positioning them ahead of other emerging thin-film technologies such as organic solar cells (OSCs), quantum-dot solar cells (QDSCs), copper indium germanium selenide (CIGS), and copper zinc tin sulfide selenide (CZTSSe) thin films~\cite{Akhtary2023}. Although OSCs, CZTSSe, and CIGS devices have reached certified efficiencies of 19.2\%, 15.07\%, and 23.6\%, respectively, single-junction PSCs have already demonstrated efficiencies exceeding 26.95\%, with tandem perovskite–silicon devices attaining certified values as high as 33.7\%~\cite{nrel_efficiency_2025}. Beyond peak efficiency, PSCs benefit from intrinsic material properties that facilitate high optical absorption, long carrier diffusion lengths, low recombination rates, and tolerance to defects~\cite{han2025perovskite}. In contrast, OSCs suffer from limited exciton diffusion lengths and require precise exciton dissociation strategies, while QDSCs are hindered by toxicity concerns and complex synthesis routes~\citep{Shilpa2023,Zhang2024}. Furthermore, PSCs exhibit bandgap tunability, allowing spectral matching and device integration flexibility, which further enhances their utility in tandem or semi-transparent applications~\citep{Shilpa2023,Afre2024}.

Despite these advantages, PSC performance remains constrained by interfacial energy-level mismatches and sub-optimal charge carrier extraction, particularly at the perovskite-hole transport layer (HTL) interface. This interface is prone to non-radiative recombination and poor carrier extraction, limiting the overall device efficiency. To address these interfacial limitations, recent strategies have focused on the development of double hole layer (DHL) architectures, which leverage synergistic combinations of inorganic and organic HTL materials. In such a configuration, an inorganic HTL such as CuO~\cite{qin2023design}, NiO\textsubscript{x}~\cite{Li2020}, CuI~\cite{Lu2021}, or Fe\textsubscript{3}O\textsubscript{4} \cite{Qureshi2023} is deposited adjacent to the perovskite absorber, followed by an organic HTL such as Spiro-OMeTAD to facilitate effective hole collection at the electrode. These bilayer structures enable the formation of a stepwise energy cascade, aligning the highest occupied molecular orbital (HOMO) and lowest unoccupied molecular orbital (LUMO) levels with the perovskite valence band and the metal work function, thereby minimizing recombination loss and enhancing charge extraction. Further modulation of interfacial energetics can be achieved via dopant incorporation or compositional engineering of HTL sublayers, which offers tunable control over work function, carrier mobility, and hole density~\cite{HEO2022101224}. 
Recently, the use of I$_2$O$_5$ as an inexpensive, eco-friendly inorganic oxidant has been proposed to oxidize Spiro-OMeTAD more effectively. This modification not only improves energy level alignment but also enhances charge mobility, reduces trap state density, and suppresses trap-assisted recombination, as demonstrated by Xu et al. \cite{xu2024spiro}.
In addition to interfacial electronic losses, ultrathin perovskite films—often desirable for reduced material consumption and enhanced mechanical flexibility—face considerable optical challenges. Specifically, they suffer from high surface reflection and inadequate light absorption across the full solar spectrum. As a result, light management strategies have become indispensable for boosting optical utilization in PSCs. Several approaches have been explored, including subwavelength surface textures \cite{Santbergen2022}, plasmonic nanostructures \cite{Mohammadi2021}, anti-reflection coatings \cite{Krajewski2023}, and spectral conversion layers \cite{Liang2021}.
Among these, nanoscale surface texturing stands out due to its superior ability to enhance light trapping without requiring deep etching, thereby preserving the structural integrity of the underlying layers. Compared to their microscale counterparts, nanostructures exhibit enhanced broadband antireflection through sub-wavelength optical interactions and can be fabricated with greater scalability. Recent studies have shown that carefully engineered nanophotonic structures can achieve optical path length enhancements that surpass the traditional Lambertian light trapping limit \cite{Luis2018}. Prominent examples include periodic gratings \cite{Dip2023NA,Ahmadi2024}, nano-pyramids \cite{Shi2015}, nanocones \cite{Peng2016}, tapered nanowires \cite{Ghosh2020}, nanospheres \cite{Elewa2021,Nowshin2023OMEx}, bow-tie \cite{Nowshin2023OC} and random textures \cite{Zhang2018}, all of which have demonstrated significant potential for improving the optical performance of solar cells.

In this paper, we propose a novel hierarchical ellipsoid-patterned methylammonium lead iodide (CH\textsubscript{3}NH\textsubscript{3}PbI\textsubscript{3} or MAPbI\textsubscript{3})-based PSC architecture that concurrently addresses interfacial band alignment challenges and optical inefficiencies through an integrated approach. At first, the structure incorporated a carefully engineered interface between CuO and Spiro-OMeTAD—two consecutive hole transport layers in the planar structure, where the valence band offset (VBO) was minimized to diminish energy barriers that usually impede hole extraction. This modification results in more efficient carrier collection and enhanced charge transport at the back contact, eventually boosting the cell's electrical performance. The introduction of a light-trapping design with a gradual modification of half-ellipsoid surface textures across the device construction was equally significant. The tailored textured features, in terms of shape and distribution, improve optical absorption by extending photon path lengths and minimizing reflecting losses, especially in the near-infrared spectrum, where traditional flat-layer cells are less effective. This approach combined electronic interface tuning and photonic structuring to enhance spectral utilization without thicker absorbers. The outcome was a highly efficient perovskite solar cell architecture with considerable implications for scalable, competitive thin-film photovoltaic technology.

\section{Design architecture and simulation methodology}
\begin{figure}[!t]
    \includegraphics[width= 1.0\textwidth]{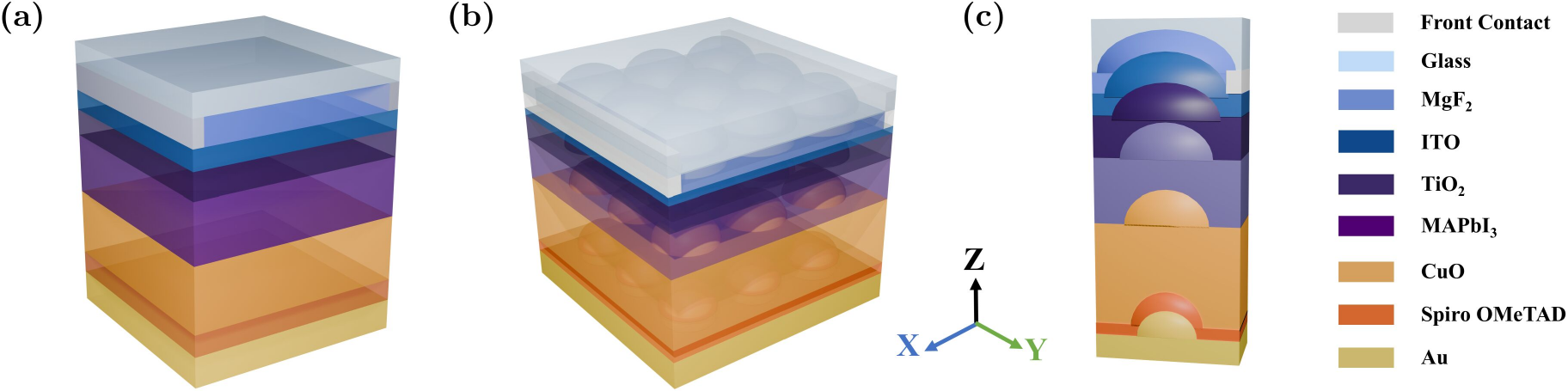}
     \caption{(a) Conventional planar double hole layer solar cell (DHLSC) architecture. (b) proposed half-ellipsoid surface solar cell (HEPSC) architecture featuring morphological optimization for improved light management and carrier extraction.}
    \label{fig:1}
\end{figure}
\begin{table}[!b]
\centering
\caption{Electrical parameters utilized for performing electrical simulations}
\label{table:1}
\resizebox{\columnwidth}{!}{%
\begin{tabular}{
>{\raggedright\arraybackslash}m{4.5cm} 
>{\centering\arraybackslash}m{1.7cm} 
>{\centering\arraybackslash}m{1.7cm} 
>{\centering\arraybackslash}m{2.2cm} 
>{\centering\arraybackslash}m{2.2cm} 
>{\centering\arraybackslash}m{2.2cm} 
>{\centering\arraybackslash}m{1cm}
}
\hline
\textbf{Parameters} & \textbf{ITO} & \textbf{TiO\textsubscript{2}} & \textbf{MAPbI\textsubscript{3}} & \textbf{CuO} & \textbf{Spiro-OMeTAD} & \textbf{Au} \\
\hline
Thickness (nm) & 50 & 10 $\sim$ 200 & 200 & 100 $\sim$ 300 & 10 $\sim$ 200 & 100 \\
DC permittivity, $\varepsilon$ & 8.9 & 9.0 & 6.5 & 18.1 & 3.0 & - \\
Bandgap, $E_g$ (eV) & 3.6 & 3.2 & 1.55 & 1.5 & 2.88, 3.02 & - \\
Electron Affinity, $\chi$ (eV) & 4.17 & 4.02 & 3.93 & 4.07 & 2.05, 2.32 & - \\
CB effective DOS $N_c$ (cm$^{-3}$) & $2.2\times10^{18}$ & $1.0\times10^{19}$ & $1.66\times10^{19}$ & $2.2\times10^{19}$ & $2.2\times10^{18}$ & - \\
VB effective DOS $N_v$ (cm$^{-3}$) & $1.8\times10^{19}$ & $1.0\times10^{19}$ & $5.41\times10^{19}$ & $5.5\times10^{20}$ & $1.8\times10^{19}$ & - \\
Mobility, $\mu$ (cm$^2$/V/s) & 40 / 30 & 20 / 10 & 50 / 50 & 10 / 0.1 & $2\times10^{-4} / 2\times10^{-4}$ & - \\
Shallow Uniform Donor Density, $N_D$ (cm$^{-3}$) & $1\times10^{21}$ & $5\times10^{18}$ & - & - & - & - \\
Shallow Uniform Acceptor Density, $N_A$ (cm$^{-3}$) & - & - & $5\times10^{13}$ & $1\times10^{15}$ & $2\times10^{18}$ & - \\
Thermal Velocity, $V_{th}$ (cm/s) & $1\times10^{7}$ & $1\times10^{7}$ & $1\times10^{7}$ & $1\times10^{7}$ & $1\times10^{7}$ & - \\
Trap-assisted SRH lifetime, $\tau_{n,p}$ (ns) & - & 5 / 2 & 8 / 8 & 100 / 100 & 
1.35 / 1.35 & - \\
Radiative Recombination Coefficient, $B_{n,p}$ (cm$^3$/s) & - & - & $2.82\times10^{-9}$ & - & - & - \\
Auger Recombination Coefficient, $A_{n,p}$ (cm$^6$/s) & - & - & $2.86\times10^{-26}$ & - & - & - \\
Defect Density, $N_t$ (cm$^{-3}$) & $5\times10^{14}$ & $1\times10^{15}$ & $2.5\times10^{15}$ & $1\times10^{15}$ & $4.93\times10^{15}$ & - \\
\hline
\end{tabular}%
}
\end{table}

Planar heterojunction perovskite solar cell structure comprising MgF\textsubscript{2}/ITO/TiO\textsubscript{2}/
MAPbI\textsubscript{3}/CuO/Spiro-OMeTAD/Au was examined initially, as illustrated in Fig.~\ref{fig:1}(a). The primary absorber layer MAPbI\textsubscript{3} is a hybrid organic-inorganic perovskite with an direct bandgap of $\sim$1.55 eV. It allowed significant light absorption while maintaining high photo-conversion efficiency in the visible range \cite{eperon2014formamidinium}. Anatase phased TiO\textsubscript{2}, as ETL with an indirect wide bandgap of 3.2 eV, had a favorable conduction band offset and preferentially conducts electrons towards the front contact ITO, while hindering hole transmission \cite{Dong2024}. Copper(II) oxide (CuO), having an indirect bandgap of 1.5 eV, was utilized to extract holes as a low-cost and stable p-type HTL, establishing a graded heterojunction with MAPbI\textsubscript{3}. Recent investigations ensured its compatibility with perovskite layers for better interfacial charge transfer \cite{Galatopoulos2017}. To enhance interface passivation and hole selectivity, Spiro-OMeTAD was incorporated as a secondary HTL. In its pristine state, Spiro-OMeTAD exhibits a wide bandgap of 2.88 eV, which increased to $\sim$3.02 eV upon oxidation, aligning well with the MAPbI\textsubscript{3} valence band and supporting efficient hole extraction \cite{xu2024spiro}. The ITO with a work function of 4.2 eV and Au electrodes with work function of 5.1 eV served as the transparent front and reflective back contacts, respectively, while the MgF\textsubscript{2} layer was implemented at the front interface as an anti-reflective coating to minimize reflection losses and improve light coupling. All interfaces in the structure—TiO\textsubscript{2}/MAPbI\textsubscript{3}, MAPbI\textsubscript{3}/CuO, and CuO/Spiro-OMeTAD—have been demonstrated experimentally in literature~\citep{Dong2024, Galatopoulos2017, Vo2024}, supporting the structural and functional feasibility of the device. 

Subsequently, the ellipsoid surface was gradually incorporated at the top of each layer surface, creating a hierarchical ellipsoid patterned solar cell (HEPSC) as shown in Figs. \ref{fig:1}(b) and (c). Ellipsoid surface optimization parameters, such as height and effective major radius of each layer, were gradually optimized from ARC to back contact. The feasibility of the fabrication was explored in a later section. Moreover, the performance metrics were evaluated under both isothermal and non-isothermal conditions to assess the thermal stability of both planar DHLSC and the proposed HEPSC.\\ 

To analyze the optical behavior of this novel device structure, the propagation of light through the different layers of the solar cell was studied using the finite-difference time-domain method (FDTD), which utilizes Maxwell's wave equations to solve the electromagnetic interaction with different layers of the device structure. The FDTD method was selected for its broad range of bands, extensive calculation capabilities, and high accuracy. The complex refractive index of materials (n+i$\kappa$) within the device structure served as the input for optical simulation. The refractive indices of ITO, TiO\textsubscript{2}, MAPbI\textsubscript{3}, CuO, Spiro-OMeTAD, and Au were obtained from the literature \citep{zhou2012direct,Tooghi2020, raoult2019optical, phillips2015dispersion, qin2023design, wen2021dynamically, olmon2012optical}. For modeling the device architecture in 2D, periodic boundary conditions were applied in the X direction and the perfectly-matched layer (PML) in the Y direction. For 3D modeling, the periodic boundary condition was applied in the X \& Y directions and PML in the Z direction. Spectral irradiance AM1.5G with the wavelength range of $\lambda \textsubscript{photon}$ from 300 nm to 1000 nm was set as the illumination source in the Y direction (2D) and the Z direction (3D) to model the optical simulation. However, in order to eliminate the effects of parasitic absorption and intra-band transitions, the photo-generation rate was evaluated exclusively within the inter-band optical transition range (300–830 nm).

For electrical stimulation, key performance parameters, such as PCE ($\eta$), short-circuit current density (J\textsubscript{sc}), open-circuit voltage (V\textsubscript{oc}), and fill factor (FF), were evaluated using the finite element method (FEM). The simulations were performed by solving Poisson’s equation, the drift–diffusion equations, and the carrier continuity equations, with Neumann boundary conditions applied at the periodic boundaries and Dirichlet boundary conditions imposed at the metal–semiconductor interfaces. 
The electrical parameters used during performing the simulation and theoretical analysis were obtained from the following literature \citep{zhou2012direct,qin2023design,kong2016simultaneous,manser2014band,ahmmed2021role,wisz2021solar,yang2024functional,ivriq2025enhancing,holzl2006work} as presented in Table \ref{table:1}. For electro-thermal analysis, parameters like mass density, specific heat, and thermal conductivity were obtained from the literature \citep{msesuppliesMgF2,chen2021mgf2,wang2018potential,bahrami2024thermal,ahmad2023impact,qian2016lattice,du2021lead,hwang2006investigation,kusiak2006cuo,battaglia2007thermophysical,dolai2017cupric,peterson2020doping, qi2020comprehensive} as shown in Table~S1 of Supplementary Material. Temperature-dependent performance was evaluated by self-consistently solving the coupled electrical and thermal transport using the finite element method, incorporating Joule heating into the heat conduction equation alongside Poisson’s, drift-diffusion, and continuity equations. Ansys Lumerical tools were used to analyze and interpret the results of the optical, thermal, and electric simulations. The simulations were performed in both 2D and 3D to refine the simulation model while simulating the device. The detailed methodology regarding optical, electrical, and thermal analysis can be found in the Supplementary Material.

\section{Results and discussion}
\subsection{Thickness and doping optimization of planar DHLSC }
Before rigorous performance analysis of the planar simulated model, the simulation parameters were considered according to the reported paper to validate the simulation results reported by Qin \textit{et al.} in 2023 \cite{qin2023design}. The final optimized efficiency of 22.77\% was closely aligned with the reported values as shown in Fig.~S1(a) of the Supplementary Material, validating our simulation process. 
However, for systematic incorporation of the hierarchical ellipsoid surface to enhance light trapping inside the perovskite absorber, the overall device thickness should be reduced. Therefore, the front contact material FTO has been replaced with ITO, and the thickness and doping levels of all layers—except the perovskite absorber—are optimized. Hence, to initialize the simulation for robust result analysis, thickness of different layers of the device has been taken ITO (100 nm), TiO\textsubscript{2} (100 nm), MAPbI\textsubscript{3} (200 nm), CuO (100 nm), Spiro-OMeTAD (100 nm), Au (100 nm), respectively, and a PCE of 19.93 \% with J\textsubscript{sc} of 23.59 mA/cm\textsuperscript{2} with V\textsubscript{oc} of 1.034 V and FF of 81.67\% was obtained as featured in Fig.~S1(d) of Supplementary Material.

\begin{figure*}[!t]
    \centering
    \includegraphics[width= 0.9\textwidth]{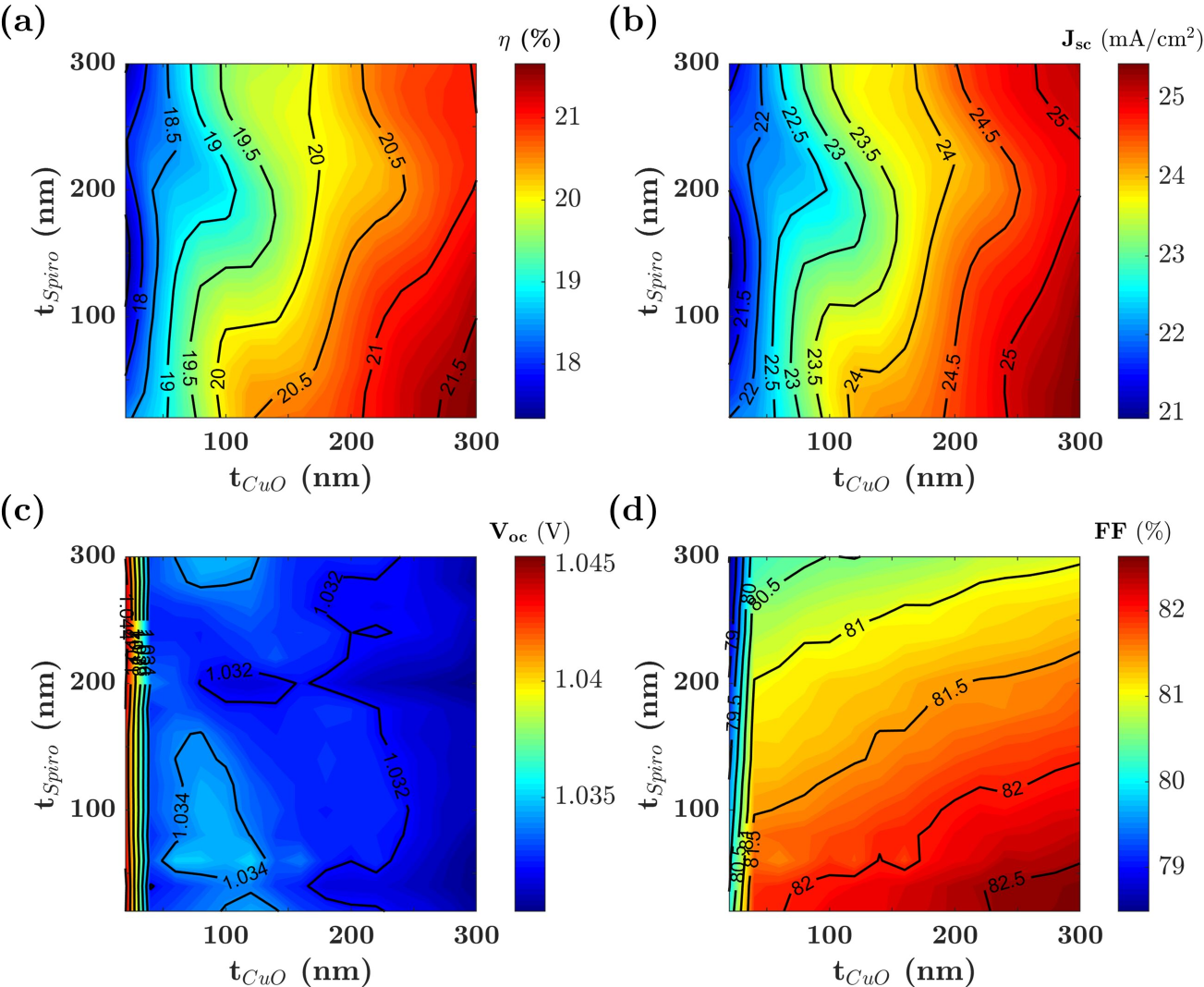}
    \caption{Impact of different structural parameters on the solar cell performance  -- (a) $\eta$ (b) J\textsubscript{sc} (c) V\textsubscript{oc} (d) FF with respect to the variation of thickness from 20  to 300 nm for CuO and Spiro-OMeTAD.}
    \label{fig:2}
\end{figure*}
\begin{figure*}[!t]
    \centering
    \includegraphics[width= 0.9\textwidth]{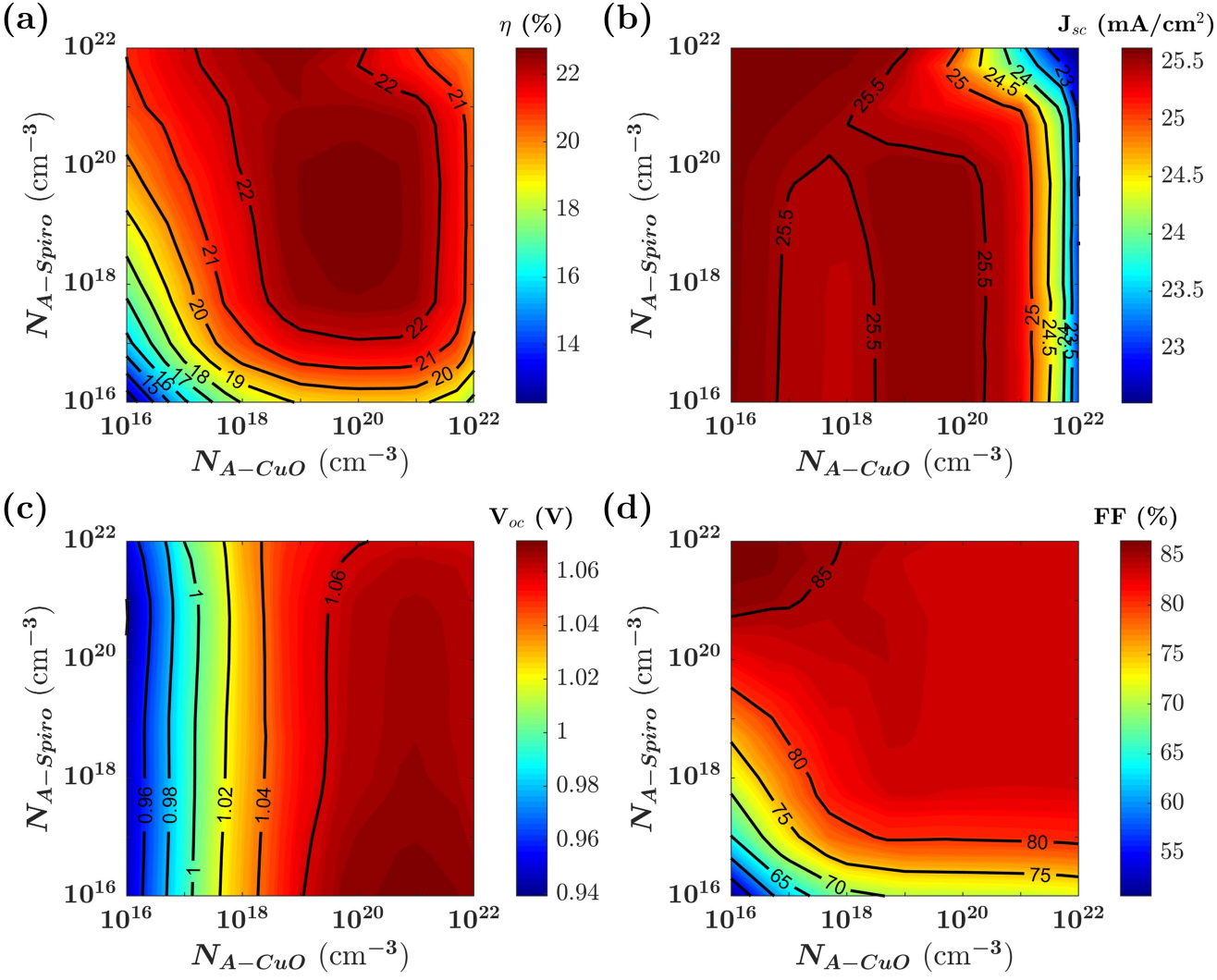}
    \caption{Impact on the performance metrics -- (a) $\eta$ (b) J\textsubscript{sc} (c) V\textsubscript{oc} (V) (d) FF for the variation of acceptor doping density ($N_{A}$) of CuO and Spiro-OMeTAD.}
    \label{fig:3}
\end{figure*}
To assess the influence of double HTL thicknesses, CuO and Spiro‑OMeTAD layers were varied from 20 to 300 nm. As shown in Fig.~\ref{fig:2}, PCE, J$_{sc}$, and FF exhibit stronger dependence on CuO due to its semiconducting nature, which enhances absorption of transmitted and reflected light, boosting carrier generation. The favorable MAPbI\textsubscript{3}/CuO band alignment supports efficient charge transport, though increasing CuO thickness slightly reduces V$_{oc}$. A peak efficiency of 21.66\% with J$_{sc}$ of 25.43 mA/cm\textsuperscript{2}, V$_{oc}$ of 1.031 V, and FF of 81.28\% was achieved at 300 nm thickness of CuO and 20 nm thickness of Spiro-OMeTAD. For stability in DHLSC configuration, a balanced structure with 300 nm CuO and 60 nm Spiro‑OMeTAD (PCE of 21.60\%) was selected for further doping optimization.

\begin{figure*}[!t]
    \centering
    \includegraphics[width= 0.85\textwidth]{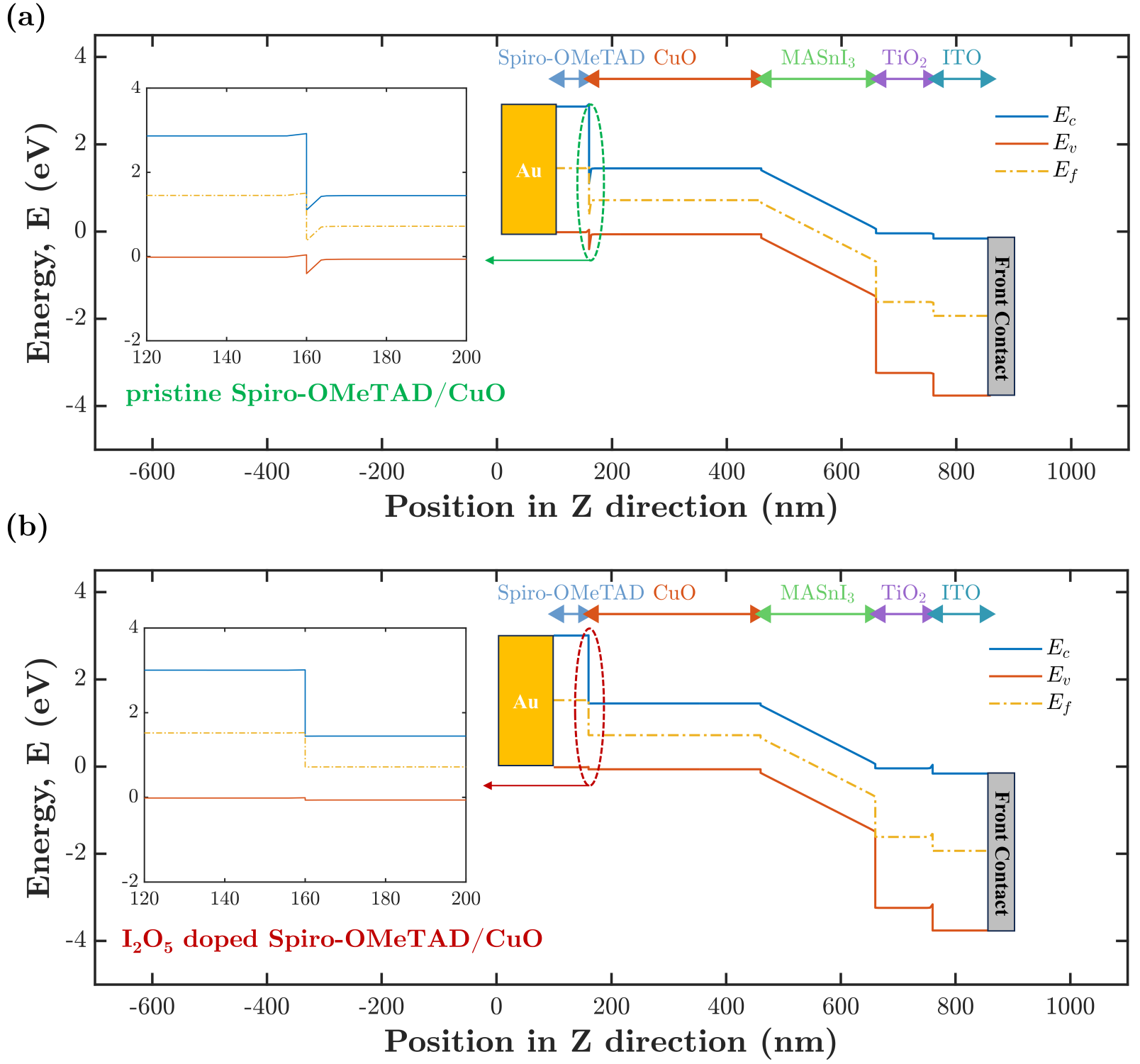}
    \caption{Band-diagram of ITO/TiO$_{2}$/MAPbI$_{3}$/CuO/Spiro-OMeTAD/Au under thermal equilibrium at short-circuit condition (V = 0V) (a) using pristine Spiro-OMeTAD (HOMO -2.05 eV, LUMO -4.93 eV) and (b) after I$_{2}$O$_{5}$ doping (HOMO -2.3 eV, LUMO -5.32 eV).}
    \label{fig:4}
\end{figure*}

The acceptor doping densities (N\textsubscript{A}) of CuO and Spiro‑OMeTAD were varied from $1 \times 10^{16}$ to $1 \times 10^{22}$ cm$^{-3}$. Enhanced PCE and J$_{sc}$ were observed within optimal doping ranges of 5$\times$10\textsuperscript{18}–-1$\times$10\textsuperscript{20} cm$^{-3}$ for CuO and 1$\times$10\textsuperscript{18}–-1$\times$10\textsuperscript{20} cm$^{-3}$ for Spiro‑OMeTAD, attributed to improved hole conductivity and stronger interfacial electric fields. A peak efficiency of 22.79\% was achieved for N$_A$ of 1$\times$10\textsuperscript{20} cm$^{-3}$ (CuO) and $1 \times 10^{19} $cm$^{-3}$ (Spiro‑OMeTAD), as shown in Fig.~\ref{fig:3} (see Table~S2 for maximum values). Beyond these levels, performance declined due to increased ionized impurity scattering and SRH recombination from defect-induced trap states. Moreover, high doping causes bandgap narrowing and energy level misalignment. To balance conductivity and recombination losses, N\textsubscript{A} of $5 \times 10^{19}$ cm$^{-3}$ (CuO) and 1$\times$10\textsuperscript{19} cm$^{-3}$ (Spiro‑OMeTAD) were selected for further analysis, offering optimal trade-offs among performance metrics.

\subsection{Band engineering of Spiro-OMeTAD HTL layer}
Following optimization of the acceptor densities of CuO and Spiro-OMeTAD HTLs, the band diagram of the whole planar solar cell was explored as shown in Fig.~\ref{fig:4}. Xu \textit{et al.} experimentally demonstrated that I$_{2}$O$_{5}$ doped Spiro-OMeTAD as HTL improved the highest occupied molecular orbitals (HOMO) level, therefore, increasing the charge carrier transportation with MAPbI$_{3}$ \cite{xu2024spiro}. Before HOMO energy level modification, there was a conduction band offset (CBO) of -0.1 eV between the lowest unoccupied molecular orbitals (LUMO) of MAPbI\textsubscript{3} and TiO\textsubscript{2} obtained from eq \ref{eq 1}. Valence band offset (VBO) of -0.55 eV  between the HOMO level of CuO and Spiro-OMeTAD extracted from the following equations,
\begin{equation}
    CBO = \chi_{\text{MAPbI\textsubscript{3}}} - \chi_{\text{TiO\textsubscript{2}}}
    \label{eq 1}
\end{equation}
\begin{equation}
    VBO = \chi_{\text{Spiro}} - \chi_{\text{CuO}} + E_{g,\text{Spiro}} - E_{g,\text{CuO}}
    \label{eq 2}
\end{equation}
By introducing I$_{2}$O$_{5}$, the conduction band energy level (LUMO) increased from -2.05 to -2.34 eV and HOMO level from -4.93 to -5.34 eV of the Spiro-OMeTAD layer. Implementing improved Spiro-OMeTAD in the DHLSC, the VBO offset in Spiro-OMeTAD/CuO junction reduced to -0.23 eV, which ensured effective transport of holes through Au contact as depicted from Fig.~\ref{fig:4}(b). After incorporating I$_2$O$_5$ doped Spiro-OMeTAD of 60 nm thickness, a PCE of 22.80\% was obtained. 
With the MAPbI\textsubscript{3} absorber fixed at 200\,nm and acceptor doping concentration of $5\times10^{13}$\,cm$^{-3}$, the effect of TiO\textsubscript{2} layer thickness (50–200\,nm) was examined under a constant donor doping of $5\times10^{18}$\,cm$^{-3}$. As shown in Fig.~S2(a) of Supplementary Material, the PCE, J$_{sc}$, and V$_{oc}$ increased with TiO\textsubscript{2} thickness up to 150 nm, beyond which performance declined due to increased series resistance introduced from trap-assisted recombination. A peak PCE of 22.89\% was achieved at 150\,nm, highlighting the critical role of ETL optimization.

Subsequently, the influence of an anti-reflective MgF\textsubscript{2} layer was investigated, resulting in a maximum PCE of 24.32\% with J\textsubscript{sc} of 27.25\,mA/cm$^2$, V\textsubscript{oc} of 1.066\,V, and FF of 83.74\% for a 100\,nm thick MgF\textsubscript{2}, as shown in Fig.~S2(b). Fig.~S3 of Supplementary Material confirmed reduced reflection from 15.4\% to 9.4\% in the 300–-830 nm wavelength range, enhancing light absorption and carrier generation, thereby boosting DHLSC performance.
\begin{figure*}[!b]
    \centering
    \includegraphics[width= 0.86\textwidth]{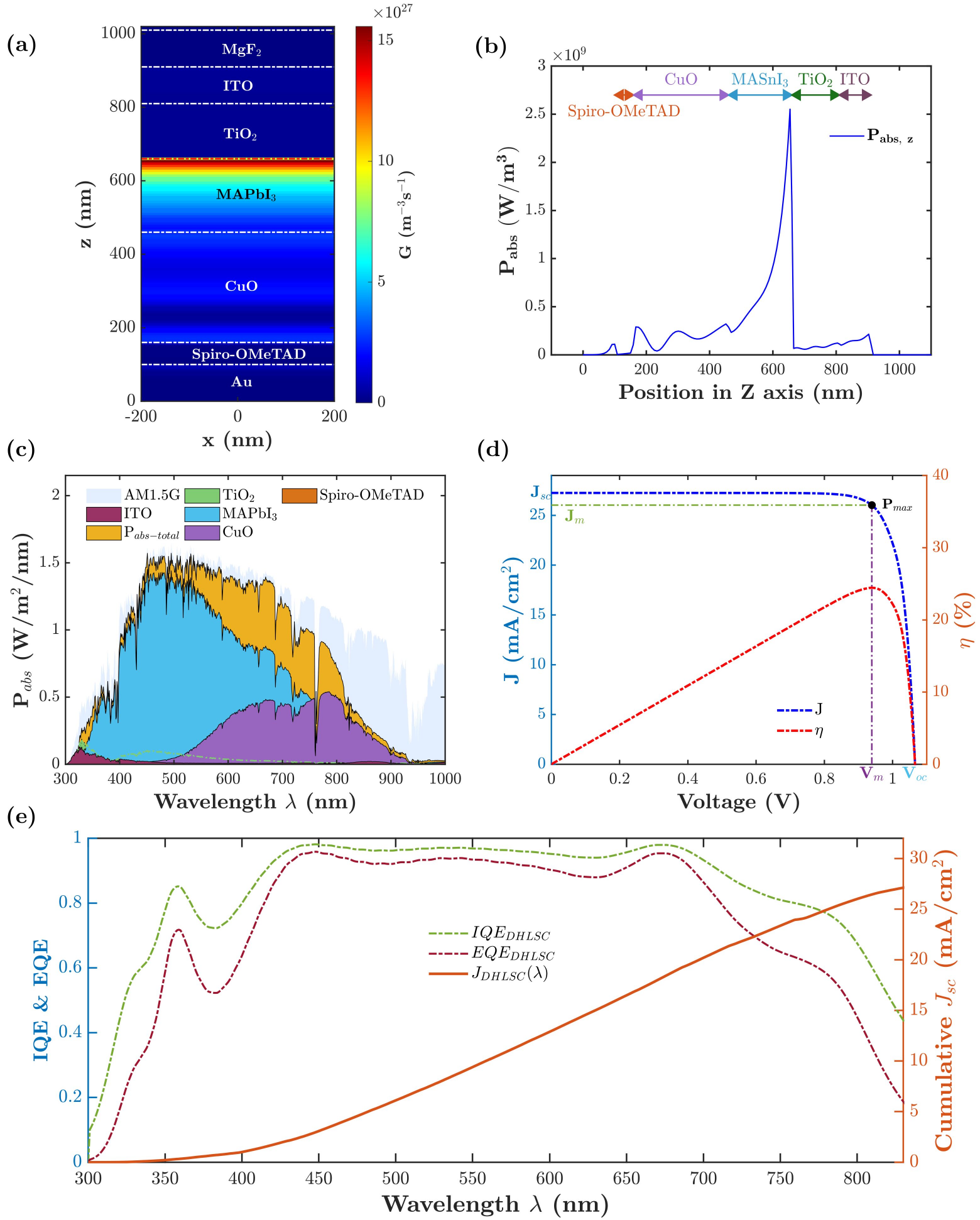}
    \caption{(a) Spatial profile of photo-generation rate, G in X-Z plane. (b) Absorber power density, P$_{abs}$, corresponding to the illumination of sunlight in the Z direction. (c) Spectral power absorption of different layers of DHLSC corresponding to the standard AM1.5G solar spectrum. (d) The J-V and $\eta$-V curves, and (e) IQE and EQE vs photon wavelength $\lambda$ featuring the performance of the optimized planar DHLSC structure.}
    \label{fig:5}
\end{figure*}

\begin{table}[!b]
\centering
\caption{Summary of performance parameters after optimization of different layers in DHLSC. }
\label{table:2}
\resizebox{\textwidth}{!}{%
\begin{tabular}{lcccc}
\hline
\multicolumn{1}{c}{Structure} &
  \begin{tabular}[c]{@{}c@{}}$\eta$\\(\%)\end{tabular} &
  \begin{tabular}[c]{@{}c@{}}J$_{sc}$ \\ (mA/cm$^{2}$)\end{tabular} &
  \begin{tabular}[c]{@{}c@{}}V$_{oc}$ \\(V)\end{tabular} &
  \begin{tabular}[c]{@{}c@{}}FF \\(\%)\end{tabular} \\ \hline
Initial Structure  & 19.93 & 23.59 & 1.034          & 81.67 \\
\begin{tabular}[c]{@{}l@{}}DHL thickness Optimization \\ (300 nm CuO, 60 nm Spiro-OMeTAD)\end{tabular}                    & 21.60 & 25.42 & 1.030          & 82.49 \\
\begin{tabular}[c]{@{}l@{}}DHL Doping Optimization\\($5\times10^{19}$ $cm^{-3}$ CuO and $1\times10^{19}$ $cm^{-3}$ Spiro-OMeTAD)\end{tabular}          & 22.78 & 25.58 & 1.063          & 83.78 \\
\begin{tabular}[c]{@{}l@{}}Energy Level matching\\(Adjusting $\chi$ and $E_{g}$ of Spiro-OMeTAD)\end{tabular} & 22.80 & 25.59 & 1.063          & 83.81 \\
\begin{tabular}[c]{@{}l@{}}TiO\textsubscript{2} thickness Optimization \\(150nm)\end{tabular}                                            & 22.90 & 25.70 & 1.063 & 83.81 \\
\begin{tabular}[c]{@{}l@{}}ARC MgF\textsubscript{2} thickness Optimization \\(100nm)\end{tabular}                                        & 24.32 & 27.25 & 1.066          & 83.74 \\ \hline
\end{tabular}%
}
\end{table}
Fig.~\ref{fig:5} presents the optoelectronic performance of the optimized double-hole-layer solar cell (DHLSC) structure. Fig. \ref{fig:5}(a) showcases the spatial distribution of the photo-generation rate G across the X-Z plane, highlighting the regions of maximum carrier generation under illumination. Fig. \ref{fig:5}(b) illustrates the absorber power density P\textsubscript{abs} along the Z-direction, indicating how effectively the structure absorbs and utilizes the incident solar energy. The power absorption of different layers of DHLSC corresponding to the standard AM1.5G solar spectrum is highlighted in Fig.~\ref{fig:5}(c). The absorption profile indicates that MAPbI$_{3}$ captures the majority of near-UV, visible, and a limited range of NIR photons within the 360–750 nm wavelength span. The secondary CuO layer, acting both as a secondary absorber and HTL layer, absorbed a significant amount of the vis-NIR spectrum of 550--830 nm.  Finally, Fig.~\ref{fig:5}(d) displays the current density–voltage (J–V) and efficiency–voltage ($\eta$–V) characteristics, capturing the electrical performance of the optimized DHLSC. In Fig.~\ref{fig:5}(e), the normalized internal quantum efficiency (IQE) and external quantum efficiency (EQE)—both unitless quantities—are plotted as functions of the photon wavelength $\lambda$, illustrating the spectral response of the device under illumination. Table~\ref{table:2} features the systematic improvement of performance with the alteration of thickness and doping concentration of different layers of planar DHLSC. The inclusion of ARC layers boosted the PCE of DHLSC from 22.90\% to 24.34\%, securing an overall improvement of 1.44\%.

\subsection{Concentrating light: introducing half ellipsoid layers}
\begin{figure*}[!b]
    \centering
    \includegraphics[width= 0.85\textwidth]{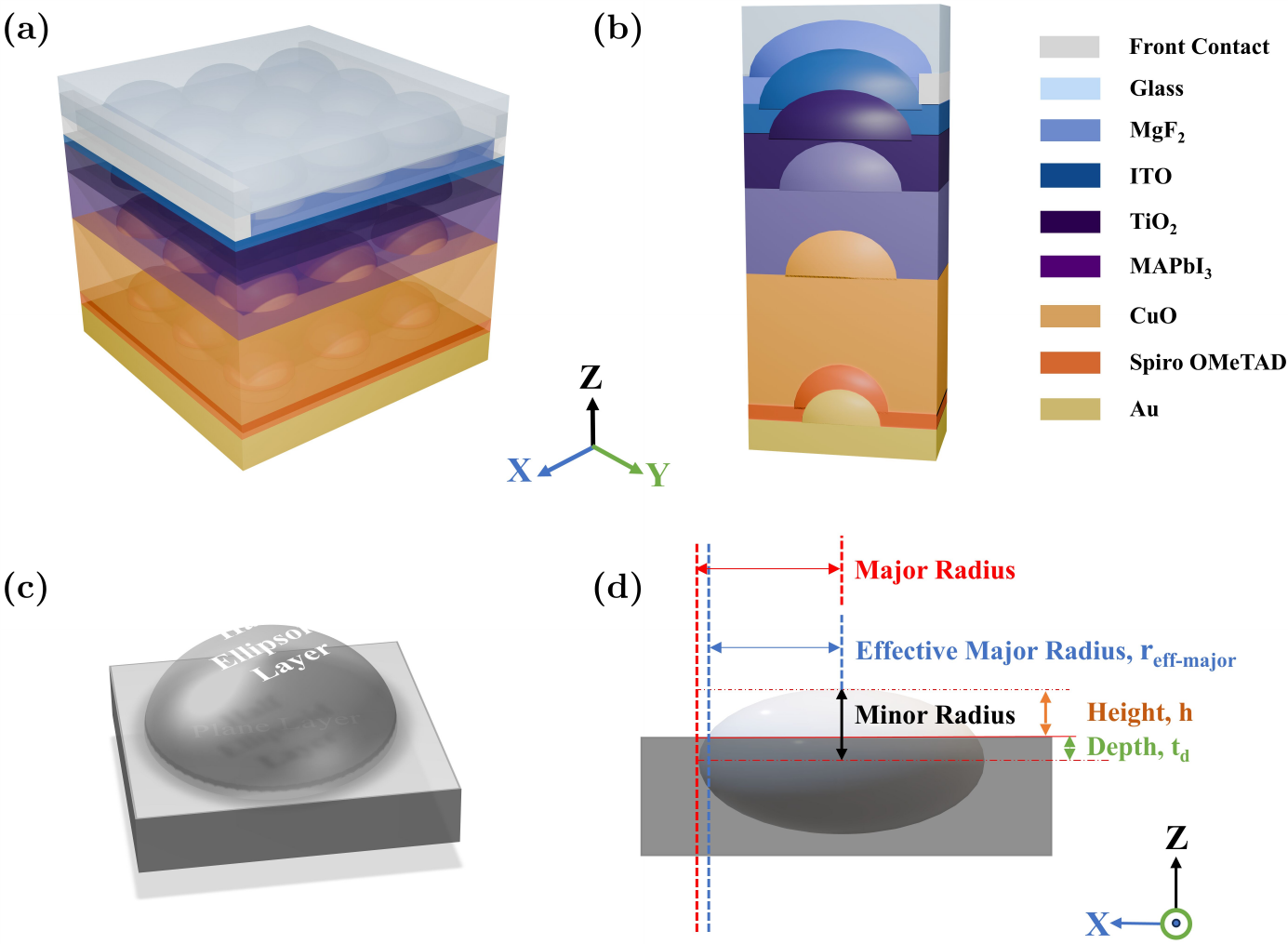}
    \caption{Schematic illustration of the proposed HEPSC architecture showing: (a) the overall 3D structure, (b) the Y–Z cross-sectional view, (c) the integration of the ellipsoidal surface atop the planar layer, and (d) the geometrical parameters considered for ellipsoidal surface optimization.}
    \label{fig:6}
\end{figure*}
\begin{figure*}[!b]
    \centering
    \includegraphics[width= 0.82\textwidth]{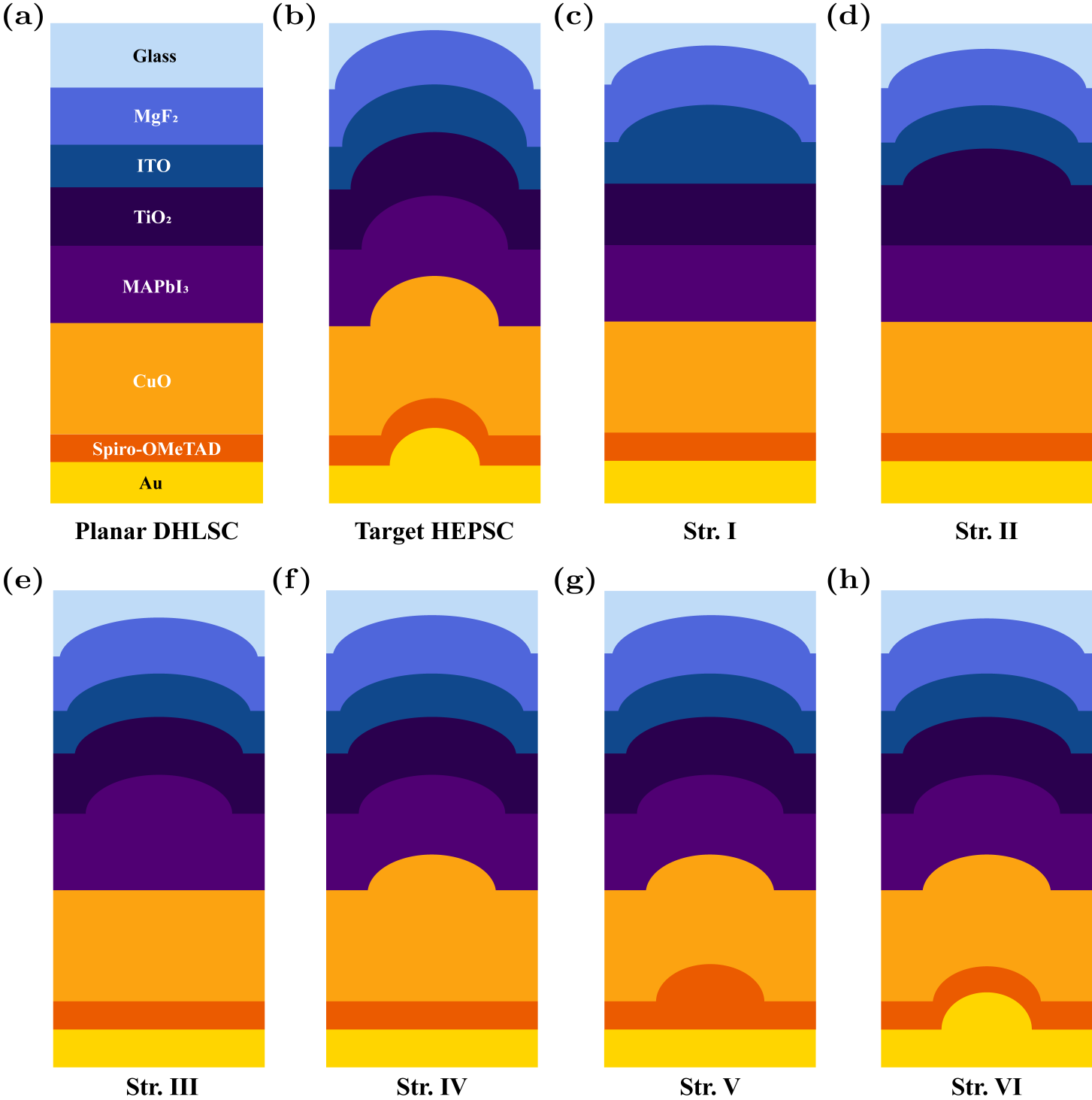}
    \caption{2D schematic representations of solar cell architectures: (a) planar DHLSC, (b) target HEPSC, and (c–h) Strs. I–VI with progressively added ellipsoidal patterned layers. In each structure, the ellipsoidal surface morphology of all previously patterned layers is preserved. The structures are labeled according to the newly added ellipsoidal layer: (c) Str. I – ITO, (d) Str. II – TiO\textsubscript{2}, (e) Str. III – MAPbI\textsubscript{3}, (f) Str. IV – CuO, (g) Str. V – Spiro-OMeTAD, and (h) Str. VI – Au.}
    \label{fig:7}
\end{figure*}

To further enhance device performance, a novel strategy was implemented by adding half-ellipsoid-shaped surfaces atop the planar layers of the optimized DHLSC cell. This hierarchical modification retains the underlying planar structure while introducing nanoscale features that strengthen photonic interactions, improve photon coupling into the absorber, and minimize reflection losses. For modeling the ellipsoid half-sphere with layers in 3D, the major axis radius (MJ) was considered for the ellipsoid radius in the X and Y axes, and the minor axis radius (MN) was considered for the radius of the ellipsoid planes in the Z axis. The major and minor axis radii are denoted as follows (MJ, MN) in nanometers. For initial study, the ellipsoidal layers (MJ, MN) radii were taken -- ITO (165 nm, 135 nm), TiO\textsubscript{2} (150 nm, 110 nm), MAPbI\textsubscript{3} (120 nm, 100 nm), CuO (110 nm, 100 nm), Spiro-OMeTAD (100 nm, 90 nm), and Au (90 nm) as a half sphere accordingly. At the top, an elliptical (185 nm, 135 nm) MgF$_2$ surface was added. After modification as shown in Fig.~\ref{fig:7}(b), an initial PCE was obtained at 25.09\% for a thickness of 300 nm for CuO and 60 nm for Spiro-OMeTAD with J\textsubscript{sc} of 28.128 mA/cm\textsuperscript{2} with V\textsubscript{oc} of 1.066 V, and FF of 83.74\%.

Although incorporating different half-ellipsoid surfaces in DHLSC provided promising efficiency, the curvature around the edge of the ellipsoid surface may not be quite practical in the case of the top layers. Therefore, to mimic a realistic model, a systematic optimization of ellipsoid surfaces above the planar layers was performed to effectively concentrate the photon energy, creating a concentrative lens-shaped photovoltaic cell. This reduces the complexity of sharp edges between different layer junctions. At first, a depth scaling factor was introduced to calculate the effective major axis radius, r\textsubscript{eff-major} in (X and Y direction) and the surface height from the following equations as featured in Fig.~\ref{fig:6}(d),
\begin{equation}
\label{eq:eff-rad-major}
    r_{\text{eff-major}} = r_{\text{major}} \times \sqrt{1 - \left( \frac{t_d}{r_{\text{minor}}} \right)^2} \\
\end{equation}
\begin{equation}
\label{eq:eff-rad-minor}
   h = r_{\text{minor}} - t_d. \\
\end{equation}
where, h denotes the surface height of ellipsoid from planar surface, $r_{\text{major}}$ denotes the major axis (MJ) radius in X and Y direction, $r_{\text{minor}}$ denotes the minor axis (MN) radius in Z direction, $t_{d}$ denotes the depth from where the ellipsoid surface is grown to create a realistic ellipsoid surface. Figs.~\ref{fig:7}(c–h) present 2D schematic representations illustrating the sequential incorporation of half-ellipsoidal surface patterning onto successive layers. Specifically, half-ellipsoidal morphologies were introduced on top of ITO (Str. I), TiO\textsubscript{2} (Str. II), MAPbI\textsubscript{3} (Str. III), CuO (Str. IV), Spiro-OMeTAD (Str. V), and Au (Str. VI), respectively. The systematic optimization of these ellipsoidal surfaces in the respective structures is described below.




\subsubsection{Influence of hierarchical photonic light trapping}
\begin{figure*}[!t]
    \centering
    \includegraphics[width= 1.0\textwidth]{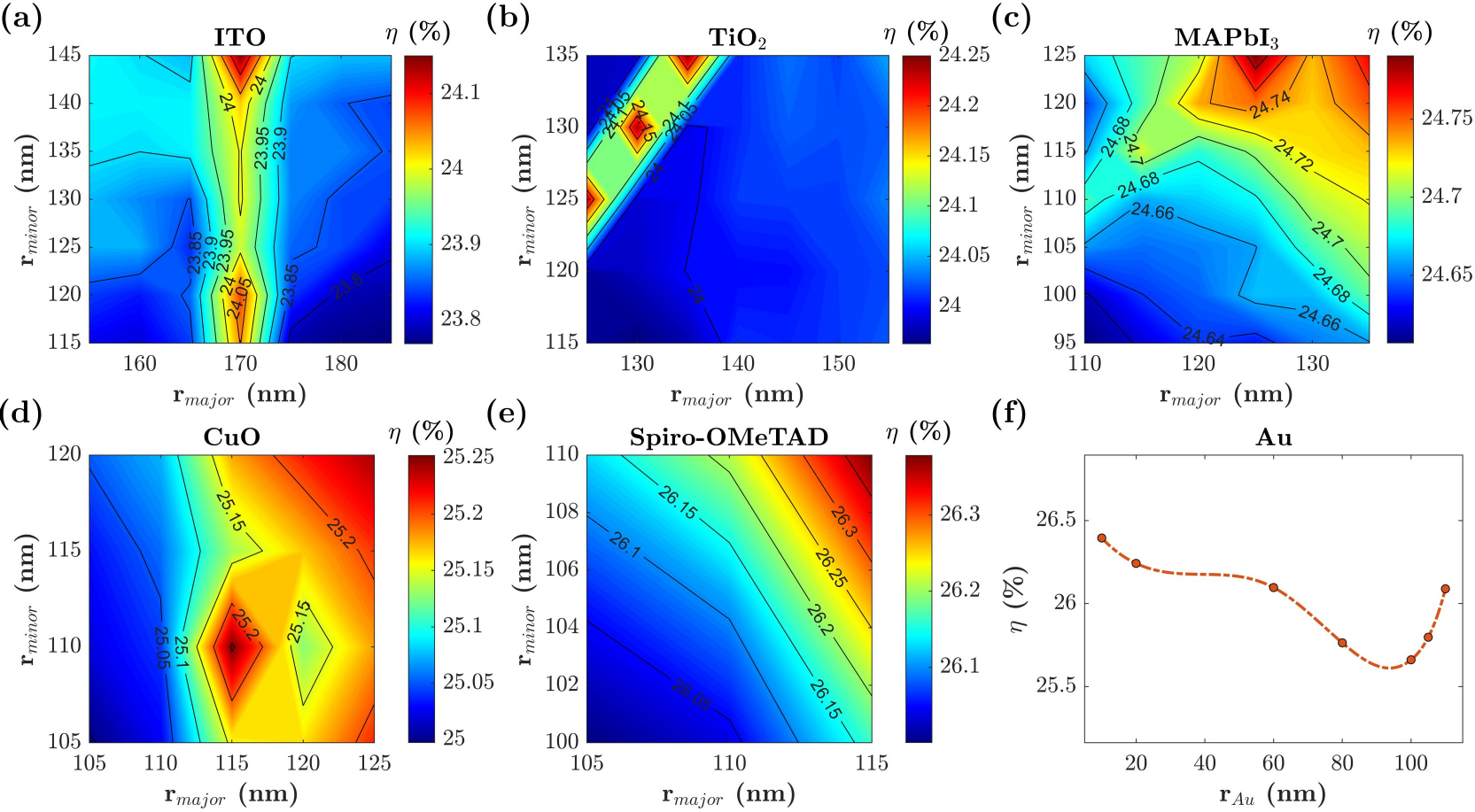}
    \caption{Impact on PCE ($\eta$) upon varying the major and minor radius of (a) ITO (b) TiO$_2$ (c) MAPbI$_3$ (d) CuO (e) Spiro-OMeTAD, and (f) radius of Au.}
    \label{fig:8}
\end{figure*}

To construct the hierarchical structure, a top–down ellipsoidal optimization strategy was employed. The optimization began with keeping a fixed top ellipsoidal MgF\textsubscript{2} ARC layer with a major radius, $r_\text{major}$ of 194 nm, and a minor radius, $r_\text{minor}$ of 140 nm, following a surface depth of $t_d$ at 10 nm as denoted in Str.~I, and illustrated in Fig.~\ref{fig:7}(c). These parameters correspond to an effective major radius, r$_{eff-major}$ of 193.5 nm and a surface height, h of 130 nm, calculated using equations~(\ref{eq:eff-rad-major}) and~(\ref{eq:eff-rad-minor}).
Subsequently, the ellipsoidal geometry of the ITO layer was varied along both the major and minor axes while the planar thickness was kept fixed at 50 nm. The $r_{major}$ was swept from 155 to 185 nm, and the $r_{minor}$ from 115 to 145 nm (surface height, $h$ of 105-135 nm). As shown in Fig.~\ref{fig:8}(a), the PCE remained relatively consistent across the full range of major radius, except at $r_{major}$ of 170 nm. For this specific major radius, a noticeable increase in PCE was observed across the full range of minor radius. This enhancement is attributed to the optimized ellipsoidal surface profile, which improves light management by trapping and redirecting incident photons laterally within the structure. The curved interfaces at this geometry increase the optical path length inside the active layers, thereby promoting enhanced carrier generation through improved photon absorption. A peak PCE of 24.15\% was achieved for a major radius of 170 nm and a minor radius of 145 nm ($h = 135$ nm).  

For Str.~II, the major and minor radius of the TiO\textsubscript{2} ETL layer were varied from 125 to 155 nm and 125 to 135 nm, respectively, while maintaining a fixed surface depth of 7 nm. The ellipsoidal surface of the ITO layer (Str. I), previously optimized, was retained above the TiO\textsubscript{2} layer. As shown in Fig.~\ref{fig:8}(b), comparable peak PCE values were observed for radius pairs (MJ, MN) of (125 nm, 125 nm), (130 nm, 130 nm), and (135 nm, 135 nm). Across most radius combinations, PCE, J$_{sc}$, and V$_{oc}$ remained consistent, as indicated by the uniform color distribution of the performance matrices for half-spherical-shaped ellipsoid surfaces as shown in Fig.~S6 of Supplementary Material. However, at the (135 nm, 135 nm) configuration, a consistent enhancement in all performance metrics is observed, with the highest PCE of 24.25\%, indicating this as the optimal radius combination. This improvement is attributed to the optimized ellipsoidal surface of the TiO\textsubscript{2} layer, which enhances forward scattering and light redirection into the underlying perovskite absorber. As a result, optical confinement is improved, leading to increased photo-generation.

In case of Str.~III,  MAPbI\textsubscript{3} absorber’s ellipsoidal surface geometry was incorporated and major and minor radius from 110–135 nm and 95–125 nm, respectively, while maintaining a fixed depth, $t_d$ of 6 nm, preserving ITO, TiO\textsubscript{2} pattern above. As illustrated in Figs.~\ref{fig:8}(c) and S7(a-d) of Supplementary Material, the device performance metrics—$\eta$, J$_{sc}$, V$_{oc}$, and FF—demonstrated a strong dependence on the surface geometry. 
Moreover, consistent enhancements in V$_{oc}$ and FF were observed when the major and minor radii were identical, indicating the importance of symmetry in ellipsoidal geometry. A peak PCE of 24.79\% was achieved at (MJ, MN) of (125 nm, 125 nm), corresponding to a surface height of 119 nm. This improvement was attributed to enhanced light confinement and trapping within the MAPbI\textsubscript{3} and CuO layers, leading to increased absorption and carrier generation.
\begin{figure*}[!b]
    \centering
    \includegraphics[width= 1.0\textwidth]{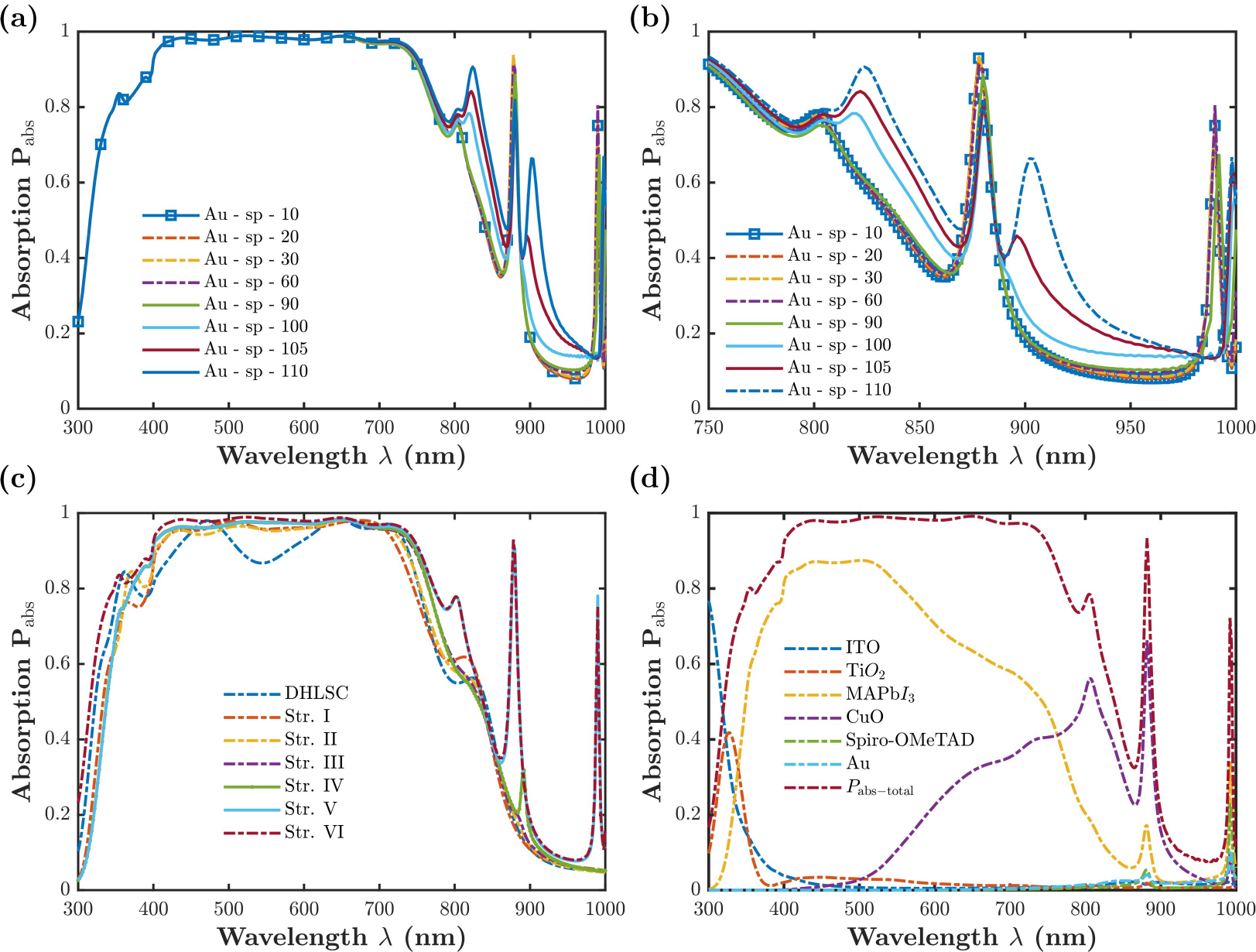}
    \caption{(a) Normalized power absorption P$_{abs}$ versus photon wavelength (300–1000 nm) for Str. VI as shown in Fig.~\ref{fig:7}(h) with varying Au sphere radius. (b) Expanded section of Fig.~\ref{fig:9}(a), highlighting the NIR region (720–1000 nm). (c) Comparison of normalized power absorption P$_{abs}$ versus photon wavelength for Str. I–VI. (d) Layer-wise contribution to the normalized power absorption of the optimized HEPSC structure (Str. V).}
    \label{fig:9}
\end{figure*}

Afterwards, CuO ellipsoidal patterned surface was introduced as Str.~IV of Fig.~\ref{fig:7} and optimized by varying its effective major and minor radius from 105 to 125 nm and 105 to 120 nm, respectively, at a fixed depth, t$_d$ of 5 nm, retaining the hierarchical pattern of top layers. A peak PCE of 25.26\% was achieved at (MJ, MN) radius of (115 nm, 110 nm), accompanied by a J$_{sc}$ of 28.08 mA/cm\textsuperscript{2}, $V_\text{oc}$ of 1.072 V, and FF of 83.88\% as shown in Figs.~\ref{fig:8}(d) and S8(a-d) of Supplementary Material.


\subsubsection{Impact of Spiro-OMeTAD ellipsoid at HTL and Au sphere at back contact}

Following Str. IV, a Spiro-OMeTAD ellipsoid surface was incorporated atop the planar layer, as illustrated in Str. V of Fig.~\ref{fig:7}(g).
The optimal device performance was achieved with major and minor radius of 115 nm and 110 nm of Spiro-OMeTAD ellipsoid surface, resulting in a peak PCE of 26.39\% as shown in Fig.~\ref{fig:8}(e). As shown in Fig.~\ref{fig:9}(c), incorporation of ellipsoidal extensions in the Spiro-OMeTAD layer, interfacing with the CuO layer respectively, led to the emergence of two pronounced absorption peaks at 880 nm and 990 nm for Str. V. This phenomenon can be explained from the electric field intensity. The comparison of the spatial distribution of the electric field intensity between Str. IV and Str. V at the X--Z cross-section of Fig. S10 revealed distinct localization. The 890 nm peak was mainly confined to the MAPbI\textsubscript{3}/CuO interface, while the 990 nm peak extended across the MAPbI\textsubscript{3}, CuO, and Spiro-OMeTAD layers.
However, the distinct peaks do not contribute to the generation of electron-hole pairs. This indicates that the ellipsoidal geometry induces layer-specific enhancement of light--matter interactions. These peaks arise from resonant optical field enhancement facilitated by the embedded ellipsoidal nanostructures. The combined morphological optimization of both the hole transport layers significantly enhanced charge extraction and light management, thereby boosting overall device efficiency.

Spherical Au inclusions with a radius ranging from 10 to 110 nm were introduced into the back contact layer while maintaining the ellipsoidal morphology above, as illustrated for Str.~VI in Fig.~\ref{fig:7}(h). 
Similarly to the optical absorption of Str. V as shown in Fig.~\ref{fig:9}(c), incorporation of Au sphere atop Au layers, interfacing with the Spiro-OMeTAD layer, led to the emergence of two pronounced absorption peaks at 880 nm and 990 nm for the Str. VI architecture. The optical absorption was predominantly governed by structural scattering and interference effects, with only minor variations in the absorption profile observed as the spherical Au radius increased, as shown in Figs.~\ref{fig:9}(a) and (b). These results indicate that, within the present multilayer architecture, the spherical Au inclusions do not play a significant role in localized surface plasmon enhancement. Instead, they function as passive scattering centers. Correspondingly, the PCE shows a decreasing trend with increasing Au sphere radius, as summarized in Table~S4 of the Supplementary Material and illustrated in Fig.\ref{fig:8}(f). This reduction in PCE arises mainly from the nano-patterning of the Au back contact, which perturbs its continuity and slightly reduces the effective interfacial area for hole extraction from the Spiro-OMeTAD layer. Additionally, the surface irregularities introduced by the Au spheres can enhance local carrier recombination and marginally increase the series resistance. Together, these effects reduce carrier collection efficiency, accounting for the modest decrease in PCE despite largely unchanged optical absorption.
\\

\begin{figure*}[!t]
    \centering
    \includegraphics[width= 1.0\textwidth]{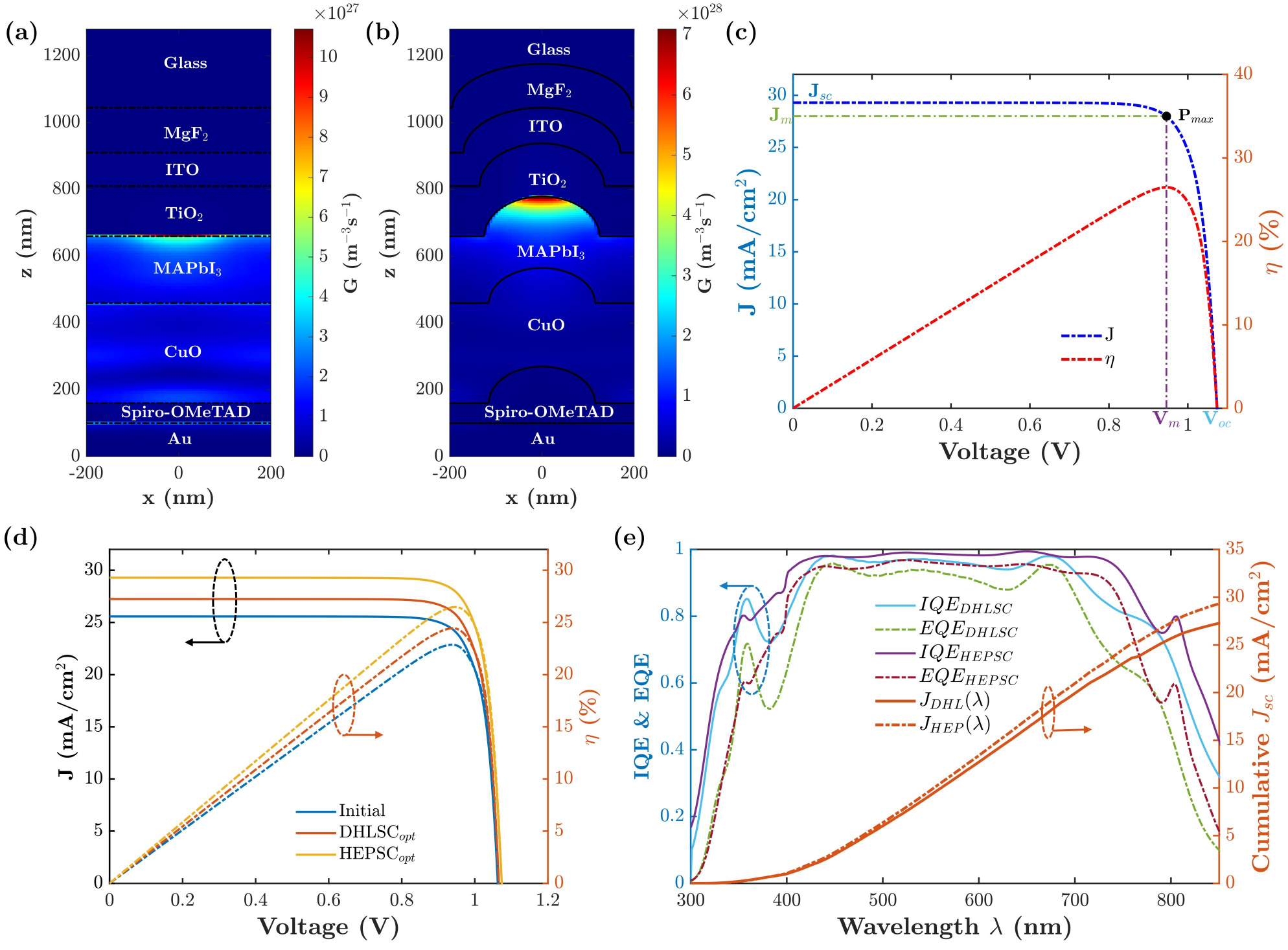}
    \caption{Spatial profile of photo-generation rate, G at (a) edge and (b) center of the periodic boundary in X-Z plane. (c) $J$--$V$ and $\eta$--$V$ curves featuring the performance of the optimized HEPSC cell. (d) $J$--$V$ and $\eta$--$V$ curves showing the gradual improvement from the initial planar to the optimized HEPSC cell. (e) Comparison of cumulative J, normalized IQE, and EQE between DHLSC and HEPSC versus photon wavelength $\lambda$ corresponding to the AM1.5G solar spectrum.}
    \label{fig:10}
\end{figure*}

Throughout the optimization of the hierarchical ellipsoidal surfaces and the spherical Au nanoparticle, comparable peak PCE values were obtained for both Str.~V and Str.~VI of Figs.~\ref{fig:7}(g) and (h). The incorporation of the Au nanosphere with a 10 nm radius, led to a negligible PCE improvement of $0.04\%$. Given the added fabrication complexity of gold nanopatterning and the insignificant performance gain of Str.~VI, Str.~V with optimized major and minor radius of (115~nm, 110~nm) of Spiro-OMeTAD, was selected as the final optimized HEPSC configuration. Fig.~\ref{fig:10} presents the optoelectronic performance of the optimized HEPSC structure. The spatial profile of photo-generation rate, G, at center X--Z cross-sections of the periodic boundary showed significant generation of electron-hole pair at the top of the MAPbI$_3$ ellipsoid surface, whereas at edge X--Z cross-section, the photo-generation rate was moderate and similar to the planar DHLSC generation as observed from Figs. \ref{fig:10}(a) and (b). The champion PCE showed 26.38\% with J$_{sc}$ of 29.29 mA/cm$^2$ with V\textsubscript{oc} of 1.074 V and FF of 83.87\% for the optimized HEPSC cell. The plot of optimized HEPSC and a comparison between the gradual improvement of performance matrices from initial DHLSC to optimized HEPSC are presented in Figs.~\ref{fig:10}(c) and (d). From relative comparison of cumulative J$_{sc}$, normalized IQE, and EQE of optimized HEPSC with optimized planar DHLSC vs. photon wavelength $\lambda$ as shown in Fig.~\ref{fig:10}(e), a significant increase of absorption in the vis-NIR spectrum is observed.

\begin{figure*}[!t]
    \centering
    \includegraphics[width= 0.9\textwidth]{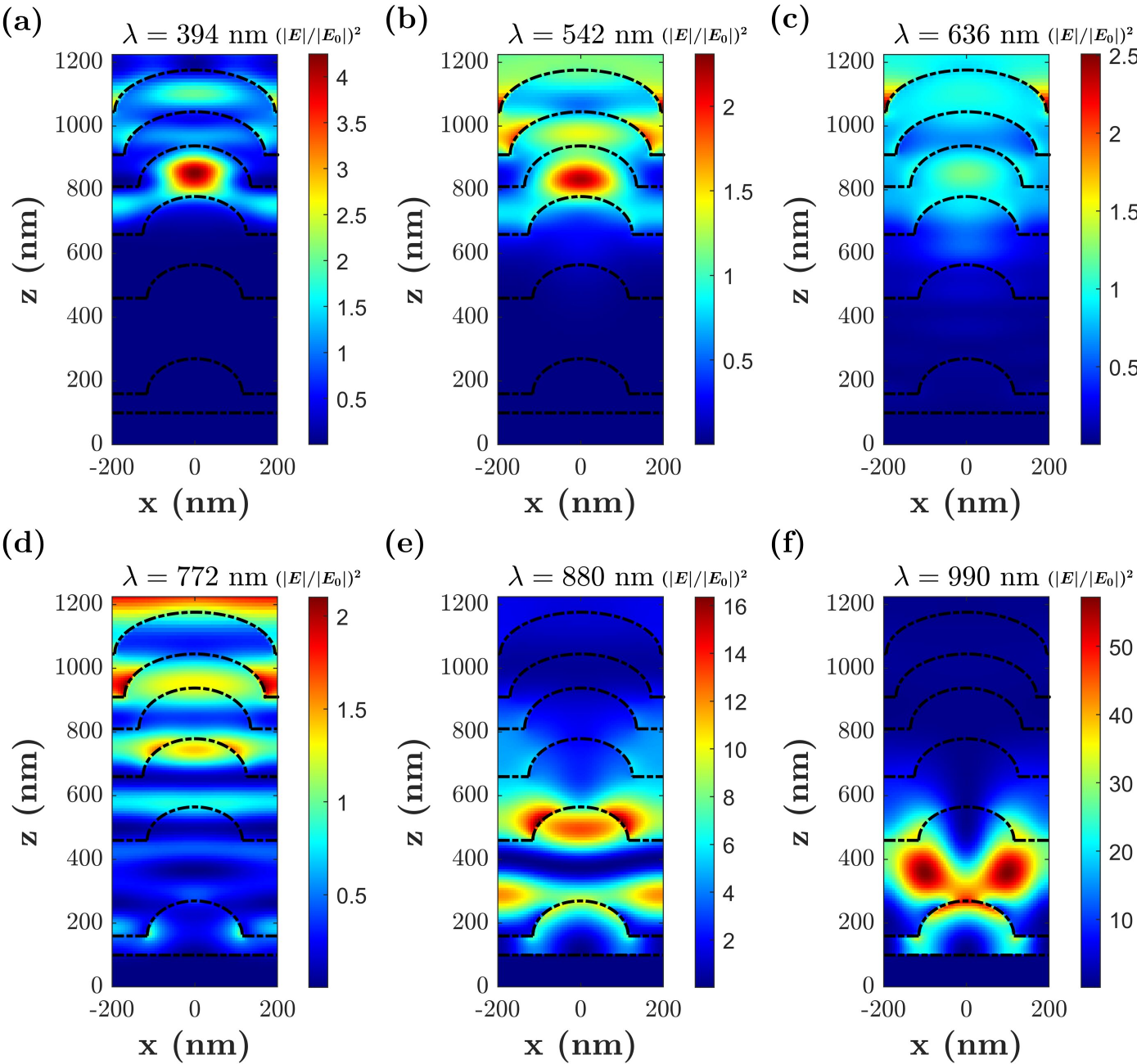}
    \caption{Spatial distribution of electric field intensity ($|\mathbf{E}|/|\mathbf{E}_0|$)\textsuperscript{2} at the center X-Z plane of the optimized HEPSC cell (Str.~V) for wavelengths $\lambda$ of (a) 394 nm, (b) 542 nm, (c) 636 nm, (d) 772 nm, (e) 880 nm, and (f) 990 nm.}
    \label{fig:11}
\end{figure*}
To analyze the spatial distribution of the electric field intensity 
$(|E|/|E_0|)^2$ inside the optimized HEPSC cell (Str. V), cross-section of electric field intensity in X-Z plane at $y=0$ corresponding to different incident wavelengths, $\lambda$ from 394 to 990 nm, are illustrated in Figs. \ref{fig:11}(a-f).
\begin{table}[!t]
\centering
\caption{Summary of performance metrics with the gradual incorporation of different half-ellipsoid surfaces above various layers}
\label{table:3}
\renewcommand{\arraystretch}{1.6}
\resizebox{\textwidth}{!}{%
\begin{tabular}{
  >{\centering\arraybackslash}m{1.9cm}  |
  >{\centering\arraybackslash}m{2.4cm}  |
  >{\centering\arraybackslash}m{5.5cm}   
  >{\centering\arraybackslash}m{1.5cm}
  >{\centering\arraybackslash}m{2.0cm}
  >{\centering\arraybackslash}m{1.5cm}
  >{\centering\arraybackslash}m{1.5cm}
  }
\hline
\textbf{Structure} & \textbf{Illustration} & \multicolumn{1}{c|}{\textbf{Parameters}} &
  \begin{tabular}[c]{@{}c@{}}$\boldsymbol{\eta}$\\(\%)\end{tabular} &
  \begin{tabular}[c]{@{}c@{}}$\boldsymbol{J_\textsc{sc}}$\\(mA/cm\textsuperscript{2})\end{tabular} &
  \begin{tabular}[c]{@{}c@{}}$\boldsymbol{V_\textsc{oc}}$\\(V)\end{tabular} &
  \begin{tabular}[c]{@{}c@{}}\textbf{FF}\\(\%)\end{tabular} \\ \hline
\textbf{Str. I} & \vspace{7 pt}\includegraphics[width=1.0cm]{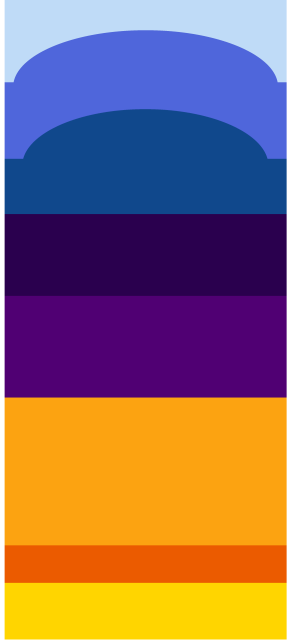} &
\multicolumn{1}{c|}{\begin{tabular}[c]{@{}c@{}}ITO ellipsoid radius\\($r_\text{major}$   = 170 nm,  $r_\text{minor}$   = 145 nm, $h$ = 135 nm)\end{tabular}} &
24.15 & 27.06 & 1.065 & 83.78 \\

\textbf{Str. II} & \includegraphics[width=1.0cm]{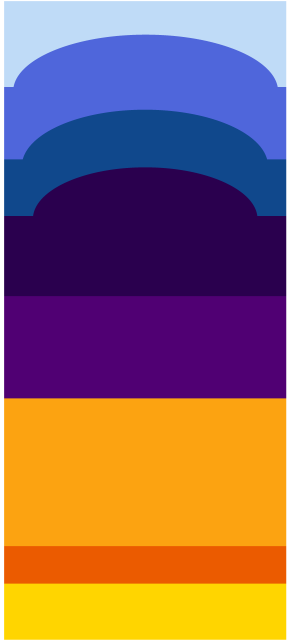} &
\multicolumn{1}{c|}{\begin{tabular}[c]{@{}c@{}}TiO\textsubscript{2} ellipsoid radius\\($r_\text{major}$ = 135 nm, $r_\text{minor}$ = 135 nm, $h$ = 128 nm)\end{tabular}} &
24.25 & 27.16 & 1.066 & 83.78 \\

\textbf{Str. III} & \includegraphics[width=1.0cm]{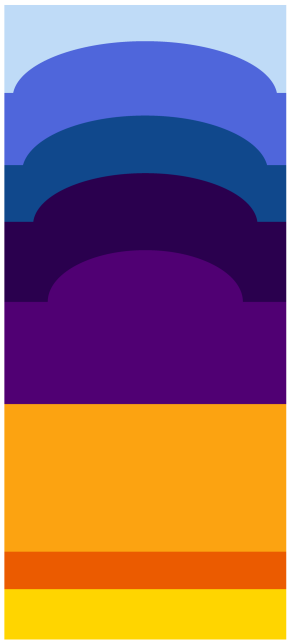} &
\multicolumn{1}{c|}{\begin{tabular}[c]{@{}c@{}}MAPbI\textsubscript{3} ellipsoid radius\\($r_\text{major}$ = 125 nm, $r_\text{minor}$ = 125 nm, $h$ = 119 nm)\end{tabular}} &
24.79 & 27.80 & 1.063 & 83.89 \\

\textbf{Str. IV} & \includegraphics[width=1.0cm]{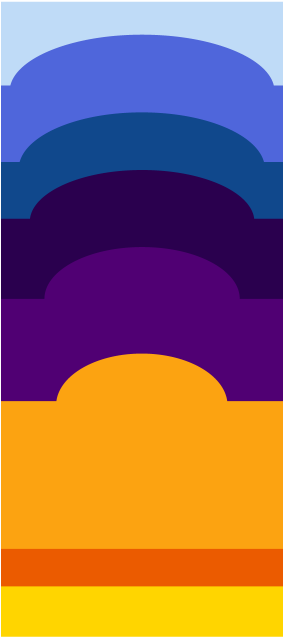} &
\multicolumn{1}{c|}{\begin{tabular}[c]{@{}c@{}}CuO ellipsoid radius\\($r_\text{major}$ = 115 nm, $r_\text{minor}$ = 110 nm, $h$ = 105 nm)\end{tabular}} &
25.26 & 28.08 & 1.072 & 83.88 \\

\textbf{Str. V} & \includegraphics[width=1.0cm]{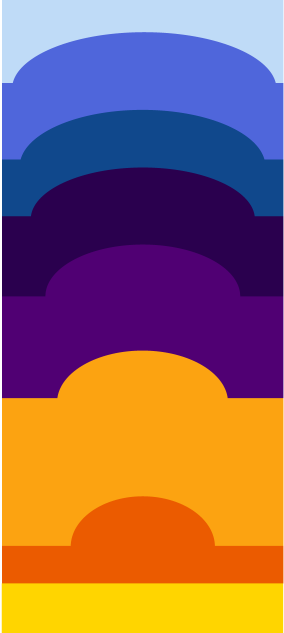} &
\multicolumn{1}{c|}{\begin{tabular}[c]{@{}c@{}}Spiro-OMeTAD ellipsoid radius\\($r_\text{major}$ = 115 nm, $r_\text{minor}$ = 110 nm, $h$ = 110 nm)\end{tabular}} &
26.38 & 29.29 & 1.074 & 83.87 \\

\textbf{Str. VI} & \includegraphics[width=1.0cm]{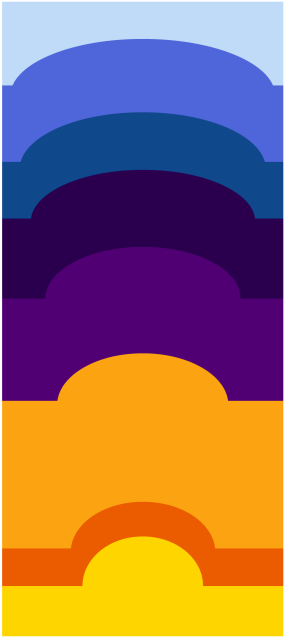} &
\multicolumn{1}{c|}{\begin{tabular}[c]{@{}c@{}}Au sphere radius\\(r =10 nm, $h$ = 10 nm)\end{tabular}} &
26.39 & 29.34 & 1.071 & 84.01 \\ \hline
\end{tabular}%
}
\end{table}
The HEPSC cell exhibited strong lateral field localization at longer wavelengths, which was enabled by the embedded ellipsoidal nanostructures. At shorter wavelengths such as 394 and 542 nm, moderate electric field enhancement was visible near the upper ellipsoids, reaching intensity values of approximately 4 and 2 times of the incident field, respectively. As the wavelength increased, more pronounced resonance effects emerged. At 772 nm, multiple ellipsoids were involved in distributed light confinement. 
At 880 nm, a strong resonant mode emerges with localized field intensities exceeding 16 times that of the incident field. At 990 nm, intense hotspots form in the lower ellipsoids and adjacent regions, with peak field intensities surpassing 50 times the incident field intensity.
The resulting spatial concentration of the E-field significantly enhances optical absorption in the active layer, particularly in the near-infrared regime, providing a clear performance advantage over the planar configuration.
A summary of the device performance of the modified HEPSC cell structure is shown in Table \ref{table:3}.

\subsection{Thermal behavior of planar DHLSC and HEPSC structure}
\begin{figure*}[!b]
    \centering
    \includegraphics[width= 0.9\textwidth]{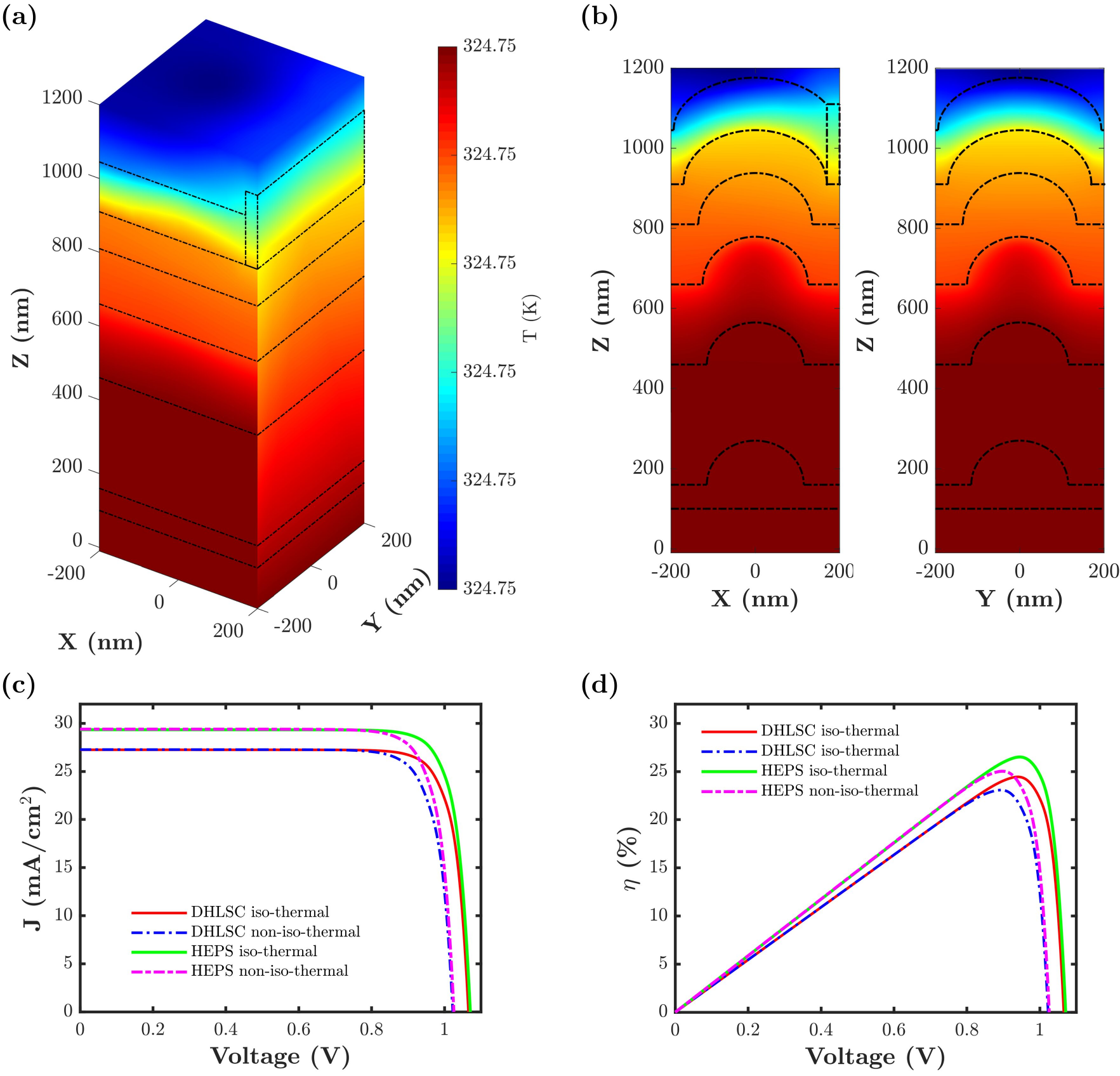}
    \caption{Thermal profile of the optimized HEPSC in (a) 3D perspective view, (b) X-Z and Y-Z center cross-section under non-isothermal steady state condition (c) J-V, and (d) $\eta$-V curve featuring the response of the planar DHLSC and optimized HEPSC in both isothermal and non-isothermal conditions.}
    \label{fig:12}
\end{figure*}
The impact of thermalization on the performance of both optimized planar DHLSC and HEPSC cells was investigated under non-isothermal steady-state conditions. The detailed methodology has been explained in Section~S1.2 of Supplementary Material. The thermal behavior and its influence on device performance were analyzed under a steady-state condition with an ambient environment temperature of 300 K (27$^\circ$C). At first, the thermal behavior of the planar DHLSC was analyzed, revealing a steady-state temperature rise from 300 K to approximately 323 K (50$^\circ$C), as shown in Fig.~S11 of Supplementary Material. Localized heating within the MAPbI\textsubscript{3}/CuO/Spiro-OMeTAD layers highlighted active carrier thermalization, exacerbated by the low thermal conductivity of Spiro-OMeTAD, which limited heat dissipation through the Au contact. As summarized in Table~\ref {table:4}, thermal effects led to a PCE drop from 24.32\% to 23.02\%, with V$_{oc}$ and FF declining from 1.066 to 1.023 V and from 83.74 to 82.54\%, respectively. A slight increase in J$_{sc}$ from 27.25 to 27.27 mA/cm\textsuperscript{2} was observed due to reduced built-in potential and enhanced carrier transport. Overall, the device retained ~94.65\% of its original efficiency under elevated temperature.

\begin{table}[!ht]
\centering
\caption{Summary of simulated performance metrics for different solar cell architectures under isothermal and non-isothermal conditions.}
\label{table:4}
\resizebox{\columnwidth}{!}{%
\begin{tabular}{lcccccc}
\hline
\textbf{Structure} & \textbf{\begin{tabular}[c]{@{}c@{}}Simulation\\ Condition\end{tabular}} & \textbf{\begin{tabular}[c]{@{}c@{}}T\\($^\circ$C)\end{tabular}} & \textbf{\begin{tabular}[c]{@{}c@{}}V$_{oc}$\\(V)\end{tabular}} & \textbf{\begin{tabular}[c]{@{}c@{}}J$_{sc}$\\(mA/cm$^2$)\end{tabular}} & \textbf{\begin{tabular}[c]{@{}c@{}}FF\\(\%)\end{tabular}} & \textbf{\begin{tabular}[c]{@{}c@{}}$\eta$\\(\%)\end{tabular}} \\
\hline
\multirow{2}{*}{\begin{tabular}[c]{@{}l@{}}Planar DHLSC\\ (With pristene Spiro-OMeTAD)\end{tabular}} & Isothermal & 27 & 1.063 & 25.58 & 83.78 & 22.78 \\
& Non-isothermal & 49 & 1.023 & 25.61 & 82.57 & 21.65 \\
\hline
\multirow{2}{*}{\begin{tabular}[c]{@{}l@{}}Planar DHLSC\\ (With I\textsubscript{2}O\textsubscript{5} doped Spiro-OMeTAD)\end{tabular}} 
& Isothermal     & 27 & 1.066 & 27.25          & 83.74 & 24.32 \\
& Non-isothermal & 50 & 1.023 & 27.27 & 82.54 & 23.02 \\
\hline
\multirow{2}{*}{\begin{tabular}[c]{@{}l@{}}Proposed HEPSC\\ (Hierarchical Ellipsoid)\end{tabular}} 
& Isothermal     & 27 & 1.074 & 29.29          & 83.87 & 26.38 \\
& Non-isothermal & 52 & 1.026 & 29.39 & 82.69 & 24.93 \\
\hline
\end{tabular}%
}
\end{table}

For the HEPSC cell, Fig.~\ref{fig:12}(a) presents the three-dimensional steady-state thermal distribution of the proposed HEPSC under non-isothermal conditions. The corresponding X–Z and Y–Z cross-sectional views in Fig.~\ref{fig:12}(b) reveal a gradual temperature gradient throughout the device, with the highest temperature localized near the absorber/HTL layers—MAPbI$_3$, CuO, and Spiro-OMeTAD—reaching approximately 325 K (52$^\circ$C). The colormap from blue to red illustrates heat accumulation, propagating vertically from the absorber toward the contact interfaces. This localized heating was primarily attributed to carrier thermalization losses and non-radiative recombination, compounded by the intrinsically low thermal conductivity of the Spiro-OMeTAD layer, which limited efficient heat flow to the metallic Au back electrode. Thermal effects on photovoltaic performance are clearly reflected in the J–V and efficiency–voltage ($\eta$–$V$) curves in Figs.~\ref{fig:12}(c–d). Despite a minor increase in J$_{sc}$ from 29.29 to 29.39 mA/cm$^2$ due to improved carrier mobility at elevated temperatures, V$_{oc}$ dropped from 1.074 to 1.026 V, and FF decreased from 83.87\% to 82.69\%. These shifts resulted in a reduction in overall PCE from 26.38\% to 24.93\% (Table~\ref{table:4}), corresponding to a retention of ~94.5\% of the original PCE under elevated thermal conditions. The decrease in $V_{oc}$ under thermal stress was due to the temperature-induced narrowing of the absorber’s effective band gap and the reduction in the built-in potential (V$_{bi}$), which lowered the energetic barrier for charge separation. Concurrently, increased thermal excitation led to enhanced recombination, especially via non-radiative pathways -- thereby compromising FF. While a modest enhancement in J$_{sc}$ was observed due to thermally assisted transport, it was insufficient to counterbalance the losses in V$_{oc}$ and FF. 
Table~\ref{table:4} highlights the thermo-electric performance of the structures from the initial to the final HEPSC cell. Despite the rise in operating temperature, the HEPSC structure retained approximately 94.5\% of its initial power conversion efficiency, demonstrating remarkable thermal resilience. This level of performance retention under non-isothermal steady-state conditions highlighted the inherent robustness of the hierarchical ellipsoidal architecture. Such stability, even in the presence of localized heating and thermal gradients, underscores the potential of this design for practical photovoltaic applications, particularly in environments where passive thermal management is limited.

\section{Comparative analysis with the existing structure}
\begin{table}[!ht]
\centering
\caption{Comparison of the performance metrics of MAPbI$_3$-based different single junction solar cells}
\label{table:5}
\resizebox{\columnwidth}{!}{%
\begin{tabular}{lccccccc}
\hline
\textbf{Structure} & \textbf{\begin{tabular}[c]{@{}c@{}}Study\end{tabular}} & 
\textbf{\begin{tabular}[c]{@{}c@{}}$\eta$\\(\%)\end{tabular}} &
\textbf{\begin{tabular}[c]{@{}c@{}}J$_{sc}$\\(mA/cm$^2$)\end{tabular}} &
\textbf{\begin{tabular}[c]{@{}c@{}}V$_{oc}$\\(V)\end{tabular}} &
\textbf{\begin{tabular}[c]{@{}c@{}}FF\\(\%)\end{tabular}} &
\textbf{\begin{tabular}[c]{@{}c@{}}Year\end{tabular}} &
\textbf{\begin{tabular}[c]{@{}c@{}}Ref\end{tabular}}\\
\hline
\begin{tabular}[c]{@{}c@{}c@{}}ITO/Cu$_2$O/MAPbI$_3$/TiO$_2$-NDs/MoSe$_2$/\\Al/SiO$_2$ \end{tabular} & Theoretical  & 19.17 & 21.91 & 1.07   & 0.82  & 2025 & \cite{Maleki2025} \\
ITO/TiO$_2$/MAPbI$_3$/CuSCN/Au                         & Theoretical  & 19.54 & 23.50 & 0.99 & 83.31 & 2020 & \cite{Tooghi2020} \\
ITO/SnO$_2$/MAPbI$_3$/Spiro-OMeTAD/Au                  & Theoretical  & 26.68 & 27.09 & 1.21 & 81.35 & 2022 & \cite{Haque2022} \\
ITO/PTAA/sc-MAPbI$_3$/C$_{60}$/BCP/Cu                    & Experimental & 21.09 & 23.46 & 1.07  & 83.50  & 2019 & \cite{Chen2019} \\
FTO/c-TiO$_2$/HC-MAPbI$_3$/Spiro-OMeTAD/Au             & Experimental & 20.22 & 24.67 & 1.10   & 74.63 & 2022 & \cite{Lu2022} \\
\begin{tabular}[c]{@{}c@{}}FTO/TiO$_2$–NPs/Cs$_{0.17}$FA$_{0.83}$Pb(I$_{0.83}$Br$_{0.17}$)$_3$/\\MAPbI$_3$ QDs/
Spiro-OMeTAD/C \end{tabular} &
  Experimental &
  16.11 &
  20.65 &
  1.08 &
  72.58 &
  2025 &
  \cite{Tipparak2025} \\
\begin{tabular}[c]{@{}c@{}c@{}}SiO$_2$/ITO/SnO$_2$/C$_6$F$_{12}$-MAPbI$_3$/\\Spiro-OMeTAD/Ag \end{tabular}     & Experimental & 21.25 & 24.21 & 1.13   & 77.68 & 2025 & \cite{Ke2025} \\
\begin{tabular}[c]{@{}c@{}c@{}}FTO/NiO$_x$/CNC:MAPbI$_3$/PC$_{61}$BM/\\BCP/Ag  \end{tabular} & Experimental & 19.90 & 24.43 & 1.07   & 76.10  & 2024 & \cite{Feng2024} \\
\textbf{\begin{tabular}[c]{@{}c@{}c@{}}MgF$_2$/ITO/TiO$_2$/MAPbI$_3$/CuO/\\I$_2$O$_5$-Spiro-OMeTAD/Au\\(DHLSC)\end{tabular}} &
  Theoretical &
  24.32 &
  27.25 &
  1.06 &
  83.74 &
  2025 &
  This Work \\
\textbf{\begin{tabular}[c]{@{}c@{}c@{}} MgF$_2$/ITO/TiO$_2$/MAPbI$_3$/CuO/\\I$_2$O$_5$-Spiro-OMeTAD/Au\\(HEPSC)\end{tabular}} &
  Theoretical &
  26.38 &
  29.29 &
  1.07 &
  83.87 &
  2025 &
  This Work \\ \hline
\end{tabular}%
}
\end{table}

Table~\ref{table:5} presents a comparative overview of performance parameters between our proposed HEPSC design and several previously reported perovskite solar cell structures. Maleki~\textit{et al.} proposed a metasurface-based inverted PSC employing TiO$_2$ nanodiscs within a MoSe$_2$ layer, which showed improved reflectivity and electron transport, resulting in a PCE of 19.17\%~\cite{Maleki2025}. Tooghi~\textit{et al.} enhanced light absorption and carrier collection by integrating convex nanostructures and a plasmonic back reflector, achieving a slightly higher PCE of 19.54\%~\cite{Tooghi2020}. However, both studies primarily focused on enhancing optical response and showed limited performance gains under thermal or structural variation. Haque \textit{et al.} applied a coupled optical-electrical modeling approach on photonic-structured ultrathin PSCs and projected significant PCE improvements, with J\textsubscript{sc} increased by over 20\% compared to planar counterparts~\cite{Haque2022}. Yet, such structures rely on ultrathin absorber layers and lack considerations for thermal stability or fabrication scalability. Meanwhile, Lu~\textit{et al.} reported that tuning MAPbI$_3$ crystal orientation from (110) to (200) via hot-casting significantly improved grain size and charge transport, yielding an efficiency of 20.22\%~\cite{Lu2022}. Recent material-level enhancements, such as the MAPbI$_3$ quantum dot-based interface layers proposed by Tipparak \textit{et al.} and the C$_6$F$_{12}$ additive-assisted film optimization by Ke \textit{et al.}, have shown improved PCEs of 16.11\% and 21.25\%, respectively~\citep{Tipparak2025, Ke2025}. These works focused primarily on improving passivation and film uniformity but remained constrained by planar device architecture and moderate current densities. Feng \textit{et al.} introduced cellulose nanocrystals (CNC)-doped MAPbI\textsubscript{3} layers to passivate defects and improve film morphology, achieving a PCE of 20.72\% with a J$_{sc}$ of 25.03 mA/cm\textsuperscript{2}~\cite{Feng2024}. Chen~\textit{et al.} employed a thick single-crystal MAPbI\textsubscript{3} absorber, achieving a PCE of 21.09\% and J$_{sc}$ of 23.46 mA/cm\textsuperscript{2}, though the fabrication complexity and scalability remain challenges~\cite{Chen2019}. By comparison, our proposed HEPSC structure outperforms all referenced structures across key photovoltaic metrics. The optimized HEPSC design achieved a PCE of 26.38\%, with J$_{sc}$ of 29.29 mA/cm$^2$, V$_{oc}$ of 1.074 V, and FF of 83.87\%, under isothermal steady-state conditions. The enhanced performance primarily stems from the use of a dual HTL stack, comprising CuO and I$_2$O$_5$-doped Spiro-OMeTAD, which promotes effective energy level alignment and hole extraction, along with the integration of a hierarchical ellipsoidal nanostructure across the device stack that significantly improves broadband light absorption from the ultraviolet to near-infrared region. Furthermore, unlike many previous works that neglected thermal behavior, our design includes a coupled opto-electro-thermal simulation, confirming that 94.5\% of peak efficiency was retained at elevated temperatures (up to 52$^\circ$C)—a critical advantage for real-world solar module deployment.
\section{Possible fabrication process of proposed structure}
To fabricate the hierarchical ellipsoid-patterned perovskite solar cell with the architecture Glass/MgF$_2$/ITO/TiO$_2$/MAPbI$_3$/CuO/Spiro-OMeTAD/Au, a combined bottom-up and top-down approach is necessary. Initially, ellipsoidal nano-patterns are etched directly into the glass substrate using laser interference lithography (LIL) or nanoimprint lithography (NIL), followed by reactive ion etching (RIE) or wet chemical etching with hydrofluoric acid (HF) for precise curvature control. Among them, NIL combined with RIE offers smoother and more uniform ellipsoidal patterns with high reproducibility and is suitable for scalable processing \cite{Lee2014}. Once the nanostructured glass is prepared, MgF$_2$ anti-reflection layer is deposited by thermal evaporation or atomic layer deposition (ALD) to ensure smooth surface coverage without altering the pattern geometry \cite{Sugai2024}. 
Etching MgF$_2$ thin films can be achieved through various methods, including physical and chemical plasma etching, ion etching, and even wet etching with specific solutions. 
Subsequently, transparent ITO films can be deposited using DC magnetron sputtering with a 90/10 wt\% indium oxide (In\textsubscript{2}O\textsubscript{3})/tin oxide (SnO\textsubscript{2}) target. Nanostructuring of the ITO layer can be performed utilizing RIE with a mixture of Ar and H$_2$ gases, enabling simultaneous physical sputtering and chemical reduction for effective etch definition \cite{Blair2024}. Next, a compact TiO$_2$ electron transport layer is deposited and annealed, serving as the scaffold for the absorber. 

To deposit MAPbI$_3$ on etched TiO$_2$ surfaces with conformal coverage and high film quality, several techniques exist. The two-step solution method -- lead iodide (PbI$_2$) spin-coating followed by methylammonium iodide (MAI) conversion offers better control than one-step methods but may struggle with non-uniformity on textured surfaces \cite{Burschka2013}. In contrast, the hybrid vapor-solution method, where PbI$_2$ is thermally evaporated and then converted with MAI, provides excellent coverage and grain uniformity even on etched TiO$_2$. 
For etched or nanostructured TiO$_2$, hybrid vapor-assisted or co-evaporation techniques are the most effective deposition strategies~\cite{Bonomi2018}. Etching MAPbI$_3$ film can be achieved by focused ion beam (FIB) lithography, where high-energy gallium ions sputter the material, or by gas-assisted FIB using XeF$_2$ or I$_2$ to improve precision and reduce damage \cite{Liu2023}. CuO$_x$ can be grown on MAPbI$_3$ perovskite layers using pulsed chemical vapor deposition (pulsed-CVD) 
enabling conformal CuO$_x$ buffer layer formation without degrading the underlying perovskite structure \cite{Eom2021}. CuO film can be etched using dilute acetic acid at low temperatures (around 35$^\circ$C), which effectively removes cupric oxide, cuprous oxide, and cupric hydroxide without damaging the underlying layers. 
After that, Spiro-OMeTAD can be deposited as HTL using ultrasonic spray coating. This technique ensures uniform coverage over both flat and curved surfaces \cite{Thornber2022}. Gold (Au) is typically deposited as the back contact using thermal evaporation or electron-beam evaporation under high vacuum. This method ensures a uniform, conductive, and adherent metal layer on the underlying HTL. 

\section{Conclusion}
The quest for high-efficiency, cost-effective, and optically sophisticated thin-film solar cells has prompted substantial investigation into nanostructured light control techniques. Multiple research studies have investigated plasmonic and resonant geometries to address the constraints of traditional planar structures, especially in boosting near-UV and NIR absorption and carrier generation. This study investigated the optical and photovoltaic performance of a planar DHLSC with a proposed advanced HEPSC. To improve performance, the band alignment mismatch between the CuO/Spiro-OMeTAD interface was reduced by incorporating I\textsubscript{2}O\textsubscript{5} doping into Spiro-OMeTAD, which ensured proper band alignment with CuO and back contact Au. Therefore, the overall PCE of the planar solar cell improved by $\sim$1.5\% after bandgap engineering and incorporation of MgF$_{2}$ ARC layer. Gradually embedding periodic half-ellipsoid surfaces above the planar layer surfaces, the light absorption inside the solar cell was significantly improved due to the confinement of the energy from photons, resulting in enhanced excitation and generation of electron-hole pairs. A champion PCE of 26.38\% with J$_{sc}$ of 29.29 mA/cm$^2$ with V$_{oc}$ of 1.074V and FF of 83.87\% for HEPSC cell was secured for the optimum HPSC cell. As the structure possesses a self-heating effect due to thermalization, a steady-state thermocouple simulation was performed, which suggested a 94.5\% PCE retained after steady state, with the device temperature increasing up to $\sim$ 325~K (52$^\circ$C). The findings of this study provide a viable pathway toward next-generation high-efficiency, thermally stable thin film perovskite solar cells by unifying advances in interfacial energetics and nanophotonic design. 


\section{CRediT authorship contribution statement}
\textbf{Md. Faiaad Rahman}: conceptualization, methodology, visualization, software, investigation, writing – original draft, and writing – review \& editing. \textbf{Arpan Sur}: conceptualization, visualization, software, investigation, writing – original draft, and writing – review \& editing. \textbf{Ahmed Zubair}: conceptualization, methodology, visualization, resources, supervision, writing – original draft, and writing – review \& editing.

\section{Data availability statement}
The data that support the findings of this study are available from the
corresponding author upon reasonable request.

\section{Declaration of competing interest}
The authors declare that they have no known competing financial interests or personal relationships that could have appeared to influence the work reported in this paper.

\section{Acknowledgments}
The authors express their sincere gratitude to the Department of Electrical and Electronic Engineering at Bangladesh University of Engineering and Technology (BUET) for providing access to the Ansys Lumerical software and the necessary technical support, including computation facilities.\\




\begin{thebibliography}{10}
\expandafter\ifx\csname url\endcsname\relax
  \def\url#1{\texttt{#1}}\fi
\expandafter\ifx\csname urlprefix\endcsname\relax\def\urlprefix{URL }\fi
\expandafter\ifx\csname href\endcsname\relax
  \def\href#1#2{#2} \def\path#1{#1}\fi

\bibitem{Liang2025}
B.~Liang, X.~Chen, X.~Wang, H.~Yuan, A.~Sun, Z.~Wang, L.~Hu, G.~Hou, Y.~Zhao, X.~Zhang, Progress in crystalline silicon heterojunction solar cells, Journal of Materials Chemistry A 13 (2025) 2441--2477.

\bibitem{andreani2019silicon}
L.~C. Andreani, A.~Bozzola, P.~Kowalczewski, M.~Liscidini, L.~Redorici, Silicon solar cells: toward the efficiency limits, Advances in Physics: X 4~(1) (2019) 1548305.

\bibitem{afroz2025perovskite}
M.~Afroz, R.~K. Ratnesh, S.~Srivastava, J.~Singh, Perovskite solar cells: Progress, challenges, and future avenues to clean energy, Solar Energy 287 (2025) 113205.

\bibitem{Akhtary2023}
N.~Akhtary, A.~Zubair, High efficiency titanium nitride bowtie nanoparticle and upconverter layer incorporated kesterite tandem solar cell, Results in Physics 55 (2023) 107148.

\bibitem{nrel_efficiency_2025}
{National Renewable Energy Laboratory}, Best research-cell efficiency chart, \url{https://www.nrel.gov/pv/cell-efficiency.html}, accessed: 2025-07-17 (2025).

\bibitem{han2025perovskite}
J.~Han, K.~Park, S.~Tan, Y.~Vaynzof, J.~Xue, E.~W.-G. Diau, M.~G. Bawendi, J.-W. Lee, I.~Jeon, Perovskite solar cells, Nature Reviews Methods Primers 5 (2025) 3.

\bibitem{Shilpa2023}
G.~Shilpa, P.~M. Kumar, D.~K. Kumar, P.~Deepthi, V.~Sadhu, A.~Sukhdev, R.~R. Kakarla, Recent advances in the development of high efficiency quantum dot sensitized solar cells ({QDSSCs}): A review, Materials Science for Energy Technologies 6 (2023) 533--546.

\bibitem{Zhang2024}
K.-N. Zhang, X.-Y. Du, L.~Yan, Y.-J. Pu, K.~Tajima, X.~Wang, X.-T. Hao, Organic photovoltaic stability: Understanding the role of engineering exciton and charge carrier dynamics from recent progress, Small Methods 8~(2) (2024) 2300397.

\bibitem{Afre2024}
R.~A. Afre, D.~Pugliese, Perovskite solar cells: A review of the latest advances in materials, fabrication techniques, and stability enhancement strategies, Micromachines 15~(2) (2024).

\bibitem{qin2023design}
F.~Qin, J.~Chen, J.~Liu, L.~Liu, C.~Tang, B.~Tang, G.~Li, L.~Zeng, H.~Li, Z.~Yi, Design of high efficiency perovskite solar cells based on inorganic and organic undoped double hole layer, Solar Energy 262 (2023) 111796.

\bibitem{Li2020}
R.~Li, P.~Wang, B.~Chen, X.~Cui, Y.~Ding, Y.~Li, D.~Zhang, Y.~Zhao, X.~Zhang, {NiO\textsubscript{x}/Spiro} hole transport bilayers for stable perovskite solar cells with efficiency exceeding 21\%, ACS Energy Letters 5~(1) (2020) 79--86.

\bibitem{Lu2021}
C.~Lu, W.~Zhang, Z.~Jiang, Y.~Zhang, C.~Ni, {CuI/Spiro-OMeTAD} double-layer hole transport layer to improve photovoltaic performance of perovskite solar cells, Coatings 11~(8) (2021).

\bibitem{Qureshi2023}
A.~A. Qureshi, E.~R. Schütz, S.~Javed, L.~Schmidt-Mende, A.~Fakharuddin, An {Fe\textsubscript{3}O\textsubscript{4}} based hole transport bilayer for efficient and stable perovskite solar cells, Energy Advances 2 (2023) 1905--1914.

\bibitem{HEO2022101224}
D.~Heo, W.~Jang, S.~Kim, Recent review of interfacial engineering for perovskite solar cells: effect of functional groups on the stability and efficiency, Materials Today Chemistry 26 (2022) 101224.

\bibitem{xu2024spiro}
J.~Xu, J.~Wu, Q.~Chen, Y.~Wang, R.~Li, X.~Chen, Z.~Lan, W.~Sun, Spiro-ometad doped with iodine pentoxide to enhance planar perovskite solar cell performance, Journal of Alloys and Compounds 970 (2024) 172749.

\bibitem{Santbergen2022}
R.~Santbergen, M.~R. Vogt, R.~Mishima, M.~Hino, H.~Uzu, D.~Adachi, K.~Yamamoto, M.~Zeman, O.~Isabella, Ray-optics study of gentle non-conformal texture morphologies for perovskite/silicon tandems, Optical Express 30~(4) (2022) 5608--5617.

\bibitem{Mohammadi2021}
M.~H. Mohammadi, M.~Eskandari, D.~Fathi, Improving the efficiency of perovskite solar cells via embedding random plasmonic nanoparticles: Optical–electrical study on device architectures, Solar Energy 221 (2021) 162--175.

\bibitem{Krajewski2023}
M.~Krajewski, A.~Callies, M.~Heydarian, M.~Heydarian, M.~Hanser, P.~S.~C. Schulze, B.~Bläsi, O.~Höhn, Roller nanoimprinted honeycomb texture as an efficient antireflective coating for perovskite solar cells, Advanced Materials Interfaces 10~(26) (2023) 2300134.

\bibitem{Liang2021}
T.~Liang, J.~Fu, M.~Li, H.~Li, Y.~Hao, W.~Ma, Application of upconversion photoluminescent materials in perovskite solar cells: opportunities and challenges, Materials Today Energy 21 (2021) 100740.

\bibitem{Luis2018}
L.~M. Pazos-Outón, T.~P. Xiao, E.~Yablonovitch, Fundamental efficiency limit of lead iodide perovskite solar cells, The Journal of Physical Chemistry Letters 9~(7) (2018) 1703--1711.

\bibitem{Dip2023NA}
P.~P. Nakti, D.~Sarker, M.~I. Tahmid, A.~Zubair, Ultra-broadband near-perfect metamaterial absorber for photovoltaic applications, Nanoscale Advances 5 (2023) 6858--6869.

\bibitem{Ahmadi2024}
H.~Ahmadi, M.~Shahrostami, N.~Manavizadeh, Perovskite half tandem v-shaped grating nanostructure solar cells: Improvement by light trapping and carrier transfer towards efficiency enhancement, Alexandria Engineering Journal 94 (2024) 80--89.

\bibitem{Shi2015}
D.~Shi, Y.~Zeng, W.~Shen, {Perovskite/c-Si} tandem solar cell with inverted nanopyramids: realizing high efficiency by controllable light trapping, Scientific Reports 5~(1) (2015) 16504.

\bibitem{Peng2016}
G.~Peng, J.~Wu, S.~Wu, X.~Xu, J.~E. Ellis, G.~Xu, A.~Star, D.~Gao, Perovskite solar cells based on bottom-fused {TiO\textsubscript{2}} nanocones, Journal of Materials Chemistry A 4 (2016) 1520--1530.

\bibitem{Ghosh2020}
A.~Ghosh, S.~S. Dipta, M.~S. Islam, Numerical analysis of light absorption in tapered nanowire based perovksite solar cell, in: 2020 11th International Conference on Electrical and Computer Engineering (ICECE), 2020, pp. 5--8.

\bibitem{Elewa2021}
S.~Elewa, B.~Yousif, M.~E.~A. Abo-Elsoud, Improving efficiency of perovskite solar cell using optimized front surface nanospheres grating, Applied Physics A 127~(11) (2021) 854.

\bibitem{Nowshin2023OMEx}
N.~Akhtary, A.~Zubair, Light trapping using a dimer of spherical nanoparticles based on titanium nitride for plasmonic solar cells, Optical Materials Express 13~(10) (2023) 2759--2774.

\bibitem{Nowshin2023OC}
N.~Akhtary, A.~Zubair, Titanium nitride based plasmonic nanoparticles for photovoltaic application, Optics Continuum 2~(7) (2023) 1701--1715.

\bibitem{Zhang2018}
H.~Zhang, M.~Kramarenko, J.~Osmond, J.~Toudert, J.~Martorell, Natural random nanotexturing of the au interface for light backscattering enhanced performance in perovskite solar cells, ACS Photonics 5~(6) (2018) 2243--2250.

\bibitem{eperon2014formamidinium}
G.~E. Eperon, S.~D. Stranks, C.~Menelaou, M.~B. Johnston, L.~M. Herz, H.~J. Snaith, Formamidinium lead trihalide: a broadly tunable perovskite for efficient planar heterojunction solar cells, Energy \& Environmental Science 7 (2014) 982--988.

\bibitem{Dong2024}
Z.~Dong, J.~Wang, J.~Men, J.~Zhang, J.~Wu, Y.~Lin, X.~Xie, J.~Wang, J.~Zhang, High-quality {TiO\textsubscript{2}} electron transport film prepared via vacuum ultraviolet illumination for {MAPbI\textsubscript{3}} perovskite solar cells, Inorganic Chemistry 63~(12) (2024) 5709--5717.

\bibitem{Galatopoulos2017}
F.~Galatopoulos, A.~Savva, I.~T. Papadas, S.~A. Choulis, The effect of hole transporting layer in charge accumulation properties of p-i-n perovskite solar cells, APL Materials 5~(7) (2017) 076102.

\bibitem{Vo2024}
T.~M. Vo, T.~M.~H. Nguyen, R.~He, C.~W. Bark, Harnessing {Cu\textsubscript{2}O}-doped 4-tert-butylpyridine in {Spiro-OMeTAD}: Study on improved performance and longevity of perovskite solar cells, ACS Omega 9~(46) (2024) 46030--46040.

\bibitem{zhou2012direct}
Y.~Zhou, J.~W. Shim, C.~Fuentes-Hernandez, A.~Sharma, K.~A. Knauer, A.~J. Giordano, S.~R. Marder, B.~Kippelen, Direct correlation between work function of indium-tin-oxide electrodes and solar cell performance influenced by ultraviolet irradiation and air exposure, Physical Chemistry Chemical Physics 14 (2012) 12014--12021.

\bibitem{Tooghi2020}
A.~Tooghi, D.~Fathi, M.~Eskandari, High-performance perovskite solar cell using photonic--plasmonic nanostructure, Scientific Reports 10~(1) (2020) 11248.

\bibitem{raoult2019optical}
E.~Raoult, R.~Bodeux, S.~Jutteau, S.~Rives, A.~Yaiche, D.~Coutancier, J.~Rousset, S.~Collin, Optical characterizations and modelling of semitransparent perovskite solar cells for tandem applications, in: 36th European Photovoltaic Solar Energy Conference and Exhibition, 2019, pp. 757--763.

\bibitem{phillips2015dispersion}
L.~J. Phillips, A.~M. Rashed, R.~E. Treharne, J.~Kay, P.~Yates, I.~Z. Mitrovic, A.~Weerakkody, S.~Hall, K.~Durose, Dispersion relation data for methylammonium lead triiodide perovskite deposited on a (100) silicon wafer using a two-step vapour-phase reaction process, Data in Brief 5 (2015) 926--928.

\bibitem{wen2021dynamically}
Z.~Wen, J.~Lu, W.~Yu, H.~Wu, H.~Xie, X.~Pan, Q.~Xu, Z.~Zhou, C.~Tan, D.~Zhou, C.~Liu, Y.~Sun, N.~Dai, J.~Hao, Dynamically reconfigurable subwavelength optical device for hydrogen sulfide gas sensing, Photonics Research 9~(10) (2021) 2060--2067.

\bibitem{olmon2012optical}
R.~L. Olmon, B.~Slovick, T.~W. Johnson, D.~Shelton, S.-H. Oh, G.~D. Boreman, M.~B. Raschke, Optical dielectric function of gold, Physical Review B 86 (2012) 235147.

\bibitem{kong2016simultaneous}
L.~Kong, G.~Liu, J.~Gong, Q.~Hu, R.~D. Schaller, P.~Dera, D.~Zhang, Z.~Liu, W.~Yang, K.~Zhu, Y.~Tang, C.~Wang, S.-H. Wei, T.~Xu, H.~kwang Mao, Simultaneous band-gap narrowing and carrier-lifetime prolongation of organic–inorganic trihalide perovskites, Proceedings of the National Academy of Sciences 113~(32) (2016) 8910--8915.

\bibitem{manser2014band}
J.~S. Manser, P.~V. Kamat, Band filling with free charge carriers in organometal halide perovskites, Nature Photonics 8~(9) (2014) 737--743.

\bibitem{ahmmed2021role}
S.~Ahmmed, A.~Aktar, A.~B.~M. Ismail, Role of a solution-processed {V\textsubscript{2}O\textsubscript{5}} hole extracting layer on the performance of cuo-zno-based solar cells, ACS Omega 6~(19) (2021) 12631--12639.

\bibitem{wisz2021solar}
G.~Wisz, P.~Sawicka-Chudy, M.~Sibiński, Z.~Starowicz, D.~Płoch, A.~Góral, M.~Bester, M.~Cholewa, J.~Woźny, A.~Sosna-Głębska, Solar cells based on copper oxide and titanium dioxide prepared by reactive direct-current magnetron sputtering, Opto-Electronics Review 29~(3) (2021) 97--105.

\bibitem{yang2024functional}
H.~Yang, Y.~Shen, G.~Xu, F.~Yang, X.~Wu, J.~Ding, H.~Chen, W.~Chen, Y.~Wu, Q.~Cheng, C.~Jin, Y.~Li, Y.~Li, Functional {Spiro-OMeTAD}-like dopant for li-ion-free hole transport layer to develop stable and efficient n-i-p perovskite solar cells, Nano Energy 119 (2024) 109033.

\bibitem{ivriq2025enhancing}
S.~B. Ivriq, M.~H. Mohammadi, R.~S. Davidsen, Enhancing photovoltaic efficiency in half-tandem {MAPbI\textsubscript{3}/ MASnI\textsubscript{3}} perovskite solar cells with triple core-shell plasmonic nanoparticles, Scientific Reports 15~(1) (2025) 1478.

\bibitem{holzl2006work}
J.~H{\"o}lzl, F.~K. Schulte, Work function of metals, Solid surface physics (2006) 1--150.

\bibitem{msesuppliesMgF2}
{MSE Supplies LLC}, Magnesium fluoride {(MgF\textsubscript{2})} crystal, \url{https://www.msesupplies.com/products/magnesium-fluoride-mgf2-crystal}, accessed: 2025-04-28 (2025).

\bibitem{chen2021mgf2}
X.~Chen, W.~Li, S.~Dou, L.~Wang, Y.~Zhao, X.~Zhang, Y.~Li, J.~Zhao, {MgF\textsubscript{2}} as abundant and environmentally friendly electrolytes for high performance electrochromic devices, Journal of Materiomics 7~(6) (2021) 1318--1323.

\bibitem{wang2018potential}
L.~Wang, J.~Wen, C.~Yang, B.~Xiong, Potential of {ITO} thin film for electrical probe memory applications, Science and Technology of Advanced Materials 19~(1) (2018) 791--801.

\bibitem{bahrami2024thermal}
M.~Bahrami, M.~Eskandari, D.~Fathi, Thermal analysis of a plasmonic perovskite solar cell: Using coupled opto-electro-thermal ({OET}) modeling, International Journal of Energy Research 2024~(1) (2024) 3921832.

\bibitem{ahmad2023impact}
A.~A. Ahmad, L.~A. Alakhras, Q.~M. Al-Bataineh, A.~Telfah, Impact of metal doping on the physical characteristics of anatase titanium dioxide ({TiO\textsubscript{2}}) films, Journal of Materials Science: Materials in Electronics 34~(20) (2023) 1552.

\bibitem{qian2016lattice}
X.~Qian, X.~Gu, R.~Yang, Lattice thermal conductivity of organic-inorganic hybrid perovskite {CH\textsubscript{3}NH\textsubscript{3}PbI\textsubscript{3}}, Applied Physics Letters 108~(6) (2016) 063902.

\bibitem{du2021lead}
X.~Du, J.~Li, G.~Niu, J.-H. Yuan, K.-H. Xue, M.~Xia, W.~Pan, X.~Yang, B.~Zhu, J.~Tang, Lead halide perovskite for efficient optoacoustic conversion and application toward high-resolution ultrasound imaging, Nature Communications 12~(1) (2021) 3348.

\bibitem{hwang2006investigation}
Y.~Hwang, Y.~Ahn, H.~Shin, C.~Lee, G.~Kim, H.~Park, J.~Lee, Investigation on characteristics of thermal conductivity enhancement of nanofluids, Current Applied Physics 6~(6) (2006) 1068--1071.

\bibitem{kusiak2006cuo}
A.~Kusiak, J.-L. Battaglia, S.~Gomez, J.-P. Manaud, Y.~Lepetitcorps, {CuO} thin films thermal conductivity and interfacial thermal resistance estimation, The European Physical Journal Applied Physics 35~(1) (2006) 17--27.

\bibitem{battaglia2007thermophysical}
J.-L. Battaglia, A.~Kusiak, Thermophysical characterization of a {CuO} thin deposit, International Journal of Thermophysics 28 (2007) 1563--1577.

\bibitem{dolai2017cupric}
S.~Dolai, R.~Dey, S.~Das, S.~Hussain, R.~Bhar, A.~Pal, Cupric oxide {(CuO)} thin films prepared by reactive d.c. magnetron sputtering technique for photovoltaic application, Journal of Alloys and Compounds 724 (2017) 456--464.

\bibitem{peterson2020doping}
K.~A. Peterson, A.~Patterson, A.~Vega-Flick, B.~Liao, M.~L. Chabinyc, Doping molecular organic semiconductors by diffusion from the vapor phase, Materials Chemistry Frontiers 4~(12) (2020) 3632--3639.

\bibitem{qi2020comprehensive}
Y.~Qi, W.~Li, S.~Liu, X.~Ma, Comprehensive design and simulation of a composite reflector for mode control and thermal management of a high-power vcsel, Journal of the Optical Society of America B 37~(11) (2020) 3487--3495.

\bibitem{Maleki2025}
J.~Maleki, M.~Shahrostami, S.~Huang, M.~Abdi-Jalebi, Efficiency boost in perovskite solar cells \textit{via} {TiO\textsubscript{2}} nanodiscs embedded in the {MoSe\textsubscript{2}} electron transport layer revealed by optoelectronic simulations, Sustainable Energy \& Fuels 9 (2025) 1797--1811.

\bibitem{Haque2022}
S.~Haque, M.~Alexandre, C.~Baretzky, D.~Rossi, F.~D. Rossi, A.~T. Vicente, F.~Brunetti, H.~Águas, R.~A.~S. Ferreira, E.~Fortunato, M.~A. der Maur, U.~Würfel, R.~Martins, M.~J. Mendes, Photonic-structured perovskite solar cells: Detailed optoelectronic analysis, ACS Photonics 9 (2022) 2408--2421.

\bibitem{Chen2019}
Z.~Chen, B.~Turedi, A.~Y. Alsalloum, C.~Yang, X.~Zheng, I.~Gereige, A.~AlSaggaf, O.~F. Mohammed, O.~M. Bakr, Single-crystal {MAPbI\textsubscript{3}} perovskite solar cells exceeding 21\% power conversion efficiency, ACS Energy Letters 4 (2019) 1258--1259.

\bibitem{Lu2022}
R.~Lu, Y.~Liu, J.~Zhang, D.~Zhao, X.~Guo, C.~Li, Highly efficient (200) oriented {MAPbI\textsubscript{3}} perovskite solar cells, Chemical Engineering Journal 433 (2022) 133845.

\bibitem{Tipparak2025}
P.~Tipparak, W.~Passatorntaschakorn, W.~Khampa, W.~Musikpan, C.~E. Usulor, C.~Bhoomanee, S.~Singh, A.~Gardchareon, A.~Ngamjarurojana, P.~Ruankham, D.~Wongratanaphisan, The impact of {MAPbI\textsubscript{3}} quantum dots on {CsFA} perovskite solar cells: Interface and hole extraction improvement, ACS Applied Energy Materials 8 (2025) 355--365.

\bibitem{Ke2025}
H.~Ke, Q.~Zhang, J.~Zhan, Y.~Zhang, S.~Zhang, M.~Zhang, P.~Zhang, Open-air-processed perfluoro(4-methylpent-2-ene)-modified {MAPbI\textsubscript{3}} solar cells actualize 21.25\% {PCE} and excellent humidity stability, Solar Energy 294 (2025) 113508.

\bibitem{Feng2024}
Y.-C. Feng, C.-E. Cai, B.-T. Liu, H.~Yang, R.-H. Lee, Cellulose nanocrystal-incorporated {MAPbI\textsubscript{3}} for inverted perovskite solar cells with enhanced efficiency and stability, ACS Applied Energy Materials 7 (2024) 12092--12102.

\bibitem{Lee2014}
S.~H. Lee, S.~Y. Lee, S.~E. Lee, H.~Lee, H.~C. Lee, Fabrication of nano-structures on glass substrate by modified nano-imprint patterning with a plasma-induced surface-oxidized {Cr} mask, Electronic Materials Letters 10 (2014) 351--355.

\bibitem{Sugai2024}
Y.~Sugai, H.~Sugata, T.~Sugawara, S.~Muhammad, J.~Hämäläinen, N.~Lamminmäki, J.~Kostamo, Optical, chemical and coverage properties of magnesium fluoride formed by atomic layer deposition, Optical Review 31 (2024) 242--246.

\bibitem{Blair2024}
S.~F.~J. Blair, J.~S. Male, C.~P. Reardon, T.~F. Krauss, Green etching of indium tin oxide metasurfaces, Optical Materials Express 14 (2024) 1924.

\bibitem{Burschka2013}
J.~Burschka, N.~Pellet, S.-J. Moon, R.~Humphry-Baker, P.~Gao, M.~K. Nazeeruddin, M.~Grätzel, Sequential deposition as a route to high-performance perovskite-sensitized solar cells, Nature 499 (2013) 316--319.

\bibitem{Bonomi2018}
S.~Bonomi, D.~Marongiu, N.~Sestu, M.~Saba, M.~Patrini, G.~Bongiovanni, L.~Malavasi, Novel physical vapor deposition approach to hybrid perovskites: Growth of {MAPbI\textsubscript{3}} thin films by rf-magnetron sputtering, Scientific Reports 8 (2018) 15388.

\bibitem{Liu2023}
Y.~Liu, F.~Li, W.~Huang, Perovskite micro-/nanoarchitecture for photonic applications, Matter 6 (2023) 3165--3219.

\bibitem{Eom2021}
T.~Eom, S.~Kim, R.~E. Agbenyeke, H.~Jung, S.~M. Shin, Y.~K. Lee, C.~G. Kim, T.-M. Chung, N.~J. Jeon, H.~H. Park, J.~Seo, Copper oxide buffer layers by pulsed-chemical vapor deposition for semitransparent perovskite solar cells, Advanced Materials Interfaces 8~(1) (2021) 2001482.

\bibitem{Thornber2022}
T.~Thornber, O.~S. Game, E.~J. Cassella, M.~E. O’Kane, J.~E. Bishop, T.~J. Routledge, T.~I. Alanazi, M.~Togay, P.~J.~M. Isherwood, L.~C. Infante-Ortega, D.~B. Hammond, J.~M. Walls, D.~G. Lidzey, Nonplanar spray-coated perovskite solar cells, ACS Applied Materials \& Interfaces 14 (2022) 37587--37594.

\end{thebibliography}

\end{document}